\documentclass[reprint, aps, prd, showpacs]{revtex4-1}

\usepackage[utf8]{inputenc}
\usepackage[T1]{fontenc}
\usepackage{amsmath,amssymb}
\usepackage{hyperref}
\usepackage{graphicx}
\usepackage{subfigure}

\begin{document}
\title{Effective dynamics of scalar perturbations in a flat Friedmann-Robertson-Walker spacetime in Loop Quantum Cosmology}
\date{\today}

\author{Mikel Fern\'andez-M\'endez}
\email{m.fernandez.m@csic.es}
\author{Guillermo A. Mena Marug\'an}
\email{mena@iem.cfmac.csic.es}
\affiliation{Instituto de Estructura de la Materia, IEM-CSIC, Serrano 121, 28006 Madrid, Spain}
\author{Javier Olmedo}
\email{jolmedo@fisica.edu.uy}
\affiliation{Instituto de F\'{i}sica, Facultad de Ciencias, Igu\'a 4225, esq.\ Mataojo, Montevideo, Uruguay}

\begin{abstract}
We study the evolution of a homogeneous and isotropic spacetime whose spatial sections have three-torus topology, coupled to a massless scalar field with small scalar perturbations within loop quantum cosmology. We consider a proposal for the effective dynamics based on a previous hybrid quantization completed by us. Consequently, we introduce a convenient gauge fixing and adopt reduced canonical variables adapted to that hybrid quantum description. Besides, we keep backreaction contributions on the {\it background} coming from terms quadratic in the perturbations in the action of the system. We carry out a numerical analysis assuming that the inhomogeneities were in a massless vacuum state at distant past (where the initial data are set). At distant future, we observe a statistical amplification of the modes amplitude in the infrared region, as well as a phase synchronization arising from quantum gravity phenomena. A description of the perturbations in terms of the Mukhanov-Sasaki gauge invariants provides the same qualitative results. Finally, we analyze some consequences of the backreaction in our effective description.
\end{abstract}

\maketitle

\section{Introduction}

The quantization of homogeneous and isotropic spacetimes has been extensively studied in loop quantum cosmology \cite{LQC} by applying the techniques developed in loop quantum gravity \cite{LQG} to this kind of symmetry reduced solutions of general relativity. The first model quantized to full completion along these lines was a flat Friedmann-Robertson-Walker (FRW) spacetime coupled to a massless scalar field. The study of its dynamics~\cite{APS, APS2} showed that the cosmological singularity is resolved and replaced by a quantum bounce (a phenomenon in loop quantum cosmology that can be traced back to the quantum properties of the geometry \cite{mmo}), and that the evolution preserves remarkably well the semiclassicality of physical states of interest \cite{recall,reca}. Several other cosmological spacetimes have also been studied in recent years, such as FRW spacetimes with a non-flat spatial topology and/or different kinds of matter content \cite{moreFLRW}, anisotropic cosmologies of various Bianchi types \cite{anis}, and even inhomogeneous cosmological models, like e.g. the case of Gowdy spacetimes~\cite{gowdy}.

Inflationary cosmologies with small inhomogeneities \cite{inflation} are currently considered in loop quantum cosmology to reach a more accurate description of the early evolution of the Universe. In particular, one of the simplest scenarios consists in a perturbed flat FRW spacetime coupled to a massive scalar field. In the unperturbed (i.e., homogeneous) case, the effective dynamics argued to arise in loop quantum cosmology \cite{effecti} has been studied in some detail \cite{sloan}, with interesting conclusions: the probability of having sufficient inflation (or, more precisely, a number of e-foldings compatible with the cosmological observations) is high enough so as to solve the fine tuning problem present in general relativity. However, the inclusion of inhomogeneities is a necessary step to treat more realistic situations. A formal, complete, and rigorous quantization of an FRW model with either flat or closed spatial sections (and straightforwardly generalizable to open sections) containing small scalar perturbations and their backreaction up to quadratic order in the action was given in Refs.~\cite{fmo,fmo1,fmo2}. In particular, those works take advantage of the uniqueness results obtained in Ref.~\cite{fmo} for the Fock quantization of the perturbations, where it was found that there is a unique canonical pair for the field that describes the perturba\-tions ---among all those linearly related by time dependent scalings of its configuration--- and a unique class of unitarily equivalent Fock quantizations for that canonical pair ---class which includes the massless representation--- such that the corresponding  vacuum state is invariant under the spatial symmetries and the dynamical evolution can be implemented as a unitary transformation. This particular choice of canonical variables allows one to eventually recover a quantum field theory with unitary dynamics in the regime in which the background behaves approximately as an \emph{effective} classical spacetime. Other recent studies \cite{AAN1,AAN2} which provide a description at the quantum level differ in the treatment of backreaction effects (this explains the inaccessibility to a symplectic structure in the total phase space of the system and to a Hamiltonian constraint in the presence of perturbations which is claimed in those studies) and in the scaling of the field which describes the perturbations. In such studies, the zero mode of the matter scalar field (providing its homogeneous component) has been identified with an internal time, and the perturbations have been shown to evolve in that time as if they propagated on a dressed background spacetime which incorporates quantum corrections. Similar conclusions about the evolution with respect to that internal time have also been obtained in Ref.~\cite{fmo2} by adopting a kind of Bohr-Oppenheimer approximation for the physical states.

Let us mention that other works on cosmological perturbations in loop quantum gravity follow a different approach. For example, from the assumption of certain form for the quantum modifications to the effective constraints, their specific expressions can be severely restricted by imposing that the modified constraint algebra must close, without constructing a quantum theory (see, e.g., Refs.~\cite{corralg, LCBG13} and the recent review~\cite{BCGM14}). From that modified algebra, one obtains corrected equations of motion for the perturbations. In particular, owing to the change in the constraint algebra (and the dynamical equations), it has been claimed that, at high densities, the universe may enter a Euclidean regime in this formalism. Similar proposals have been put forward starting with a lattice approach in Ref.~\cite{Wilson-Ewing12}. In contrast, the effective equations that we employ here are inferred from a genuine quantum model, namely, the one described in Ref.~\cite{fmo2}. As a consequence, there is no signature change in our analysis, and the equations of motion of the perturbations remain hyperbolic even during the bounce.

Thus, in this work, and following the ideas of Refs.~\cite{bmp1,bmp} for Gowdy cosmologies, we investigate the cosmological model discussed in Ref.~\cite{fmo2}, i.e., a FRW spacetime with flat and compact spatial sections of three-torus topology coupled to a scalar field with small scalar perturbations, by assuming an effective dynamics that generalizes that of homogeneous and isotropic loop quantum cosmology to this case with inhomogeneities. In particular, our purpose is to understand how the perturbations are affected when they evolve from the distant past through the quantum bounce to the asymptotic future, and how the backreaction produced by the inhomogeneities may affect the background evolution (namely, the dynamics of the homogeneous degrees of freedom). In this sense, we can say that our work complements and extends the studies of Refs.~\cite{AAN1,AAN2}, where the initial data are chosen at the bounce (and the backreaction is neglected). For the sake of simplicity, we study the effective dynamics of this model in the case in which the field is massless. We integrate numerically the Hamilton equations of motion, sampling the phase space by considering for the inhomogeneities random initial data peaked around a massless vacuum state (up to quantum fluctuations that are modeled by certain probability distributions), and some particular trajectories for the variables describing the part of the background. In this way, we observe that the amplitudes of the modes are statistically amplified in the process of crossing the bounce, especially in the infrared sector. The amplification follows certain oscillatory patterns and decays in the ultraviolet limit. Moreover, we also observe that the phases of certain modes with the same dynamics are considerably synchronized in the expanding branch of the Universe, after the bounce: hence, if these phases were considered alone, isolated from the rest of the system, there would be a partial loss of the information that existed previous to the bounce. The very same qualitative results are obtained if one adopts a description in terms of gauge invariant variables. We interpret these phenomena as quantum gravity effects caused by the interaction at leading order between the quantum background and the quantum inhomogeneities. Finally, we analyze the deviation of the background evolution for the homogeneous degrees of freedom on some particular trajectories in which the backreaction can be considered non-negligible but still small enough. We find, in particular, that the growth of the homogeneous physical volume after the bounce is modified by the presence of perturbations.

The rest of this paper is organized as follows. In Sec.~\ref{sec:class_sys} we present the classical description of the model. The effective dynamics is detailed in Sec.~\ref{sec:semiclass_sys}. In Sec.~\ref{sec:numerics} we summarize our numerical results. Finally, the discussion and conclusions can be found in Sec.~\ref{sec:conclu}. We also have added three appendices.

\section{Classical model}\label{sec:class_sys}

The cosmological system that we will study is a flat FRW model with a matter content consisting of a massless scalar field, with both the geometry and the matter field deformed with small inhomogeneities that are treated perturbatively.

Let us start with a spacetime that admits a global time function $t$ and a foliation in spatial hypersurfaces that are topologically equivalent to a three-torus. The spacetime metric can be codified in the lapse function $N$, the shift vector $N_i$, and the spatial metric $h_{ij}$ induced on the spatial sections, where the Latin indices $i,j=1,2,3$ denote spatial indices. We are interested in situations with small inhomogeneities that can be handled as perturbations around homogeneous and isotropic solutions. Hence, the metric functions can be expressed in the following form:
\begin{subequations}\label{eq:gen-metric}
\begin{align}
& N(t,\vec\theta)=\sigma N_0(t)\big[1+{}^{(1)}\!N(t,\vec\theta)\big],\\
& N_i= \sigma^2e^{\alpha(t)} \;{}^{(1)}\!N_i(t,\vec\theta), \\
& h_{ij}(t,\vec\theta) = \sigma^2 e^{2\alpha(t)}\big[{}^{(0)}\!h_{ij}(\vec\theta)+{}^{(1)}\!h_{ij}(t,\vec\theta)\big],
\end{align}
\end{subequations}
where $N_0$ is the homogeneous lapse function, $\alpha$ is the logarithm of the scale factor, ${}^{(0)}\!h_{ij}$ is the standard flat metric on the three-torus $T^3$ with respect to the angular coordinates $\theta_i\in (0, l_0]$ (the fiducial volume of the three-torus is then $l_0^3$), and $\sigma^2=4\pi G/(3l_0^3)$, $G$ being the Newton constant. Regarding the inhomogeneous part of these metric functions, we consider the Fourier mode decomposition
\begin{subequations}\label{eq:pert-metric}
\begin{align}
& {}^{(1)}\!h_{ij}(t,\vec\theta) = 2\sum_{\vec n,\epsilon}a_{\vec n,\epsilon} (t){}^{(0)}\!h_{ij}(\vec\theta)\tilde Q_{\vec n,\epsilon}(\vec\theta)\\\nonumber
& \quad +6\sum_{\vec n,\epsilon}b_{\vec n,\epsilon}(t)\left[\frac1{\omega_n ^2}(\tilde Q_{\vec n,\epsilon})_{|ij}(\vec\theta)+\frac13{}^{(0)}\!h_{ij}(\vec\theta)\tilde Q_{\vec n,\epsilon}(\vec\theta)\right], \\
& {}^{(1)}\!N(t,\vec\theta) = \sum_{\vec n,\epsilon}g_{\vec n,\epsilon}(t)\tilde Q_{\vec n,\epsilon}(\vec\theta), \\
& {}^{(1)}\!N_i(t,\vec\theta) = \sum_{\vec n,\epsilon}\frac1{\omega_n^2}k_{\vec n,\epsilon}(t)(\tilde Q_{\vec n,\epsilon})_{|i}(\vec\theta),
\end{align}
\end{subequations}
in terms of the real Fourier harmonics $\tilde Q_{\vec n,\epsilon}$ [namely, the sine and cosine eigenfunctions of the Laplace-Beltrami operator associated to ${}^{(0)}\!h_{ij}$; see Eq. \eqref{Qharmo}], and their (covariant) derivatives $(\tilde Q_{\vec n,\epsilon})_{|i}$ and $(\tilde Q_{\vec n,\epsilon})_{|ij}$, with $\epsilon=+,-$,  $\vec n=(n_1,n_2,n_3)\in\mathbb Z^3$, and $\omega_n^2=4\pi^2\vec n\cdot\vec n /l_0^{2}$. Furthermore, summation over the tuples $\vec n$ excludes the zero mode $\vec n=(0,0,0)$, s
ince it is contained in the homogeneous part of our functions and variables. Besides, one has to restrict the summation to, e.g., tuples whose first non-vanishing component is strictly positive in order to avoid a double counting of modes (since a flip of sign in the tuple changes the harmonic at most in a global sign). For additional details see Appendix~\ref{ap:harm}.

Regarding the matter content, we will consider a massless scalar field $\Phi$. We adopt the mode decomposition
\begin{equation}\label{eq:gen-matter}
\Phi(t,\vec\theta) = \frac1{l_0^{3/2}\sigma}\left[\varphi(t)+\sum_{\vec n,\epsilon}f_{\vec n,\epsilon}(t)\tilde Q_{\vec n,\epsilon} (\vec\theta)\right],
\end{equation}
where the variable $\varphi$ is (up to a constant factor) the zero mode of the field, excluded again from the summation. Thus,
the inhomogeneities are codified by the time dependent coefficients $a_{\vec n,\epsilon}$, $b_{\vec n,\epsilon}$,
$g_{\vec n,\epsilon}$, $k_{\vec n,\epsilon}$, and $f_{\vec n,\epsilon}$.

After substituting expressions~\eqref{eq:gen-metric}, \eqref{eq:pert-metric}, and
\eqref{eq:gen-matter} in the Einstein-Hilbert action, and truncating at second order in the perturbative coefficients, we obtain a Hamiltonian that is a linear combination of first class constraints: a global Hamiltonian constraint, whose Lagrange multiplier is the homogeneous lapse and which is the sum of the constraint of the unperturbed model plus (spatially integrated) terms that are quadratic in the perturbations and their momenta, and two local constraints that are linear in the inhomogeneities, whose corresponding Lagrange multipliers are the inhomogeneous part of the lapse function and the shift vector (see Ref.~\cite{fmo}).

We now introduce a gauge fixing in order to get rid of non-physical degrees of freedom. In particular, we choose the longitudinal gauge, which is imposed by requiring the conditions (for all tuples $\vec n$) \cite{fmo}
\begin{equation}\label{eqs:gaugeB}
\pi_{a_{\vec n,\epsilon}}-\pi_\alpha a_{\vec n,\epsilon}-3\pi_\varphi f_{\vec n,\epsilon} = 0,\quad b_{\vec n,\epsilon} = 0,
\end{equation}
where each $\pi_q$ denotes the momentum canonically conjugate to the configuration variable
$q$. The imposition of the linear constraints amounts to the conditions
\begin{equation}\label{eq:a_n}
a_{\vec n,\epsilon} = 3\frac{\pi_\varphi\pi_{f_{\vec n,\epsilon}}-3\pi_\alpha\pi_\varphi f_{\vec n,\epsilon}}{9\pi_\varphi^2+\omega_n^2e^{4\alpha}}
\end{equation}
and to the vanishing of the canonical momenta of the perturbations $b_{\vec n,\epsilon}$.

As was noticed in Ref.~\cite{fmo}, this gauge fixing affects the symplectic structure by changing Poisson to Dirac brackets in a non-trivial way. However, one can introduce the following new set of canonical coordinates~\cite{fmo2}:
\begin{subequations}\label{eqs:newvariablesB}
\begin{align}
\bar f_{\vec n,\epsilon} &= e^\alpha f_{\vec n,\epsilon}, \\
\pi_{\bar f_{\vec n,\epsilon}} &= e^{-\alpha}(\pi_{f_{\vec n,\epsilon}}-3\pi_\varphi a_{\vec n,\epsilon}-\pi_\alpha f_{\vec n,\epsilon}), \\
\bar\alpha &= \alpha+\frac12\sum_{\vec n,\epsilon}\big(a_{\vec n,\epsilon}^2+f_{\vec n,\epsilon}^2\big), \\
\pi_{\bar\alpha} &= \pi_\alpha-\sum_{\vec n,\epsilon}\big(f_{\vec n,\epsilon}\pi_{f_{\vec n,\epsilon}}-3\pi_\varphi a_{\vec n,\epsilon}f_{\vec n,\epsilon}-\pi_\alpha f_{\vec n,\epsilon}^2\big), \\
\bar\varphi &= \varphi+3\sum_{\vec n,\epsilon}a_{\vec n,\epsilon}f_{\vec n,\epsilon}, \\
\pi_{\bar\varphi} &= \pi_\varphi.
\end{align}
\end{subequations}
This transformation diagonalizes the reduced symplectic structure and provides a field description adapted to the hybrid quantization put forward in Refs.~\cite{fmo,fmo1,fmo2}. This field description is achieved by scaling the configuration of the perturbations with the background scale factor, together with the inverse scaling of the momentum, which also gets a contribution linear in the field so as to remove the dominant part of the cross terms that couple the field configuration and its momentum in the Hamiltonian. Thus, in these new variables, the part of the Hamiltonian which is quadratic in the perturbations adopts a Klein-Gordon form (with corrections of the order of $\omega_n^{-2}$). Besides, the canonical transformation performed on the perturbations respects the linearity of the field equations in the inhomogeneous sector and is local (i.e., the same for all non-zero modes). It introduces quadratic corrections in the zero modes which only affect the form of the (quadratic) perturbative contribution to the Hamiltonian. Let us comment that this choice of field variables is the only one for which the quantum field dynamics can be unitarily implemented in the (effective) classical background while respecting the invariance under its spatial isometries, and that this unitary implementation can be achieved, e.g., by adopting the massless Fock representation \cite{fmo}. Classically, this choice has no relevant physical consequences (all canonical transformations lead to classically equivalent descriptions), but it is the natural one in order to study the evolution of the perturbations admitting the assumption that they were initially on a physical state of the corresponding quantization, e.g. the vacuum. Besides, this simplifies the dynamics that we want to analyze.

In the reduced system, we are left with the global Hamiltonian constraint
\begin{equation}\label{eq:reducedH}
H(N_0) = N_0\Big(H_{|0}+\sum_{\vec n,\epsilon}H^{\vec n,\epsilon}_{|2}\Big),
\end{equation}
where
\begin{equation}
H_{|0}=\frac{1}{2}e^{-3\bar \alpha}\big(-\pi_{\bar \alpha}^2+\pi_{\bar \varphi}^2\big)
\end{equation}
is the unperturbed scalar constraint, and the quadratic perturbative contribution is given by
\begin{equation}\label{eq:reducedH^n_|2}
H^{\vec n,\epsilon}_{|2} = \frac12e^{-\bar \alpha}\big(\bar E^n_{\pi\pi}\pi_{\bar f_{\vec n,\epsilon}}^2+2\bar E^n_{f\pi}\bar f_{\vec n,\epsilon}\pi_{\bar f_{\vec n,\epsilon}}+\bar E^n_{ff}\bar f_{\vec n,\epsilon}^2\big),
\end{equation}
with coefficients
\begin{subequations}
\begin{align}
\bar E^n_{\pi\pi} &= 1-\frac3{\omega_n^2}e^{-4\bar\alpha}\pi_{\bar\varphi}^2, \\
\bar E^n_{f\pi} &= \frac6{\omega_n^2}e^{-6{\bar\alpha}}\pi_{\bar\alpha}\pi_{\bar\varphi}^2, \\
\bar E^n_{ff} &= \omega_n^2-\frac12e^{-4\bar\alpha}\big(\pi_{\bar\alpha}^2+15\pi_{\bar\varphi}^2\big) -\frac{12}{\omega_n^2}e^{-8\bar\alpha}\pi_{\bar\alpha}^2\pi_{\bar\varphi}^2.
\end{align}
\end{subequations}

\section{Effective loop quantum cosmology}\label{sec:semiclass_sys}

In order to adapt the previous description to the improved dynamics scheme of loop quantum cosmology \cite{APS}, we introduce the variables
\begin{subequations}\label{eq:lqc-variables}
\begin{equation}
|v| = l_0^3\sigma^3e^{3\bar\alpha},\quad v\beta = -\gamma l_0^3\sigma^2\pi_{\bar \alpha},
\end{equation}
where $v$ is the physical volume of the Universe and $\beta$ is its canonically conjugate variable (up to a numerical factor). As for the
matter content, we introduce
\begin{equation}
\phi = \frac{\bar\varphi}{l_0^{3/2}\sigma},\quad \pi_\phi = l_0^{3/2}\sigma\pi_{\bar\varphi} ,
\end{equation}
for the parametrization of the zero modes, and
\begin{equation}
\quad \tilde f_{\vec n,\epsilon}=\frac{\bar f_{\vec n,\epsilon}}{l_0^{1/2}}, \quad \tilde \pi_{\tilde f_{\vec n,\epsilon}}=l_0^{1/2}\pi_{\bar f_{\vec n,\epsilon}}
\end{equation}
\end{subequations}
for the rest. These phase space variables satisfy the classical algebra
\begin{equation}
\{\beta,v\}=4\pi G\gamma,\ \{\phi,\pi_\phi\}=1,\ \{\tilde f_{\vec n,\epsilon},\tilde \pi_{\tilde f_{\vec n',\epsilon'}}\}=\delta_{\vec n,\vec n'}\delta_{\epsilon,\epsilon'},
\end{equation}
the other Poisson brackets being equal to zero.

The last ingredient to construct the (plausible) effective dynamics of the quantum model introduced in Refs.~\cite{fmo1,fmo2} comes from the polymerization scheme adopted in those works in order to represent integer powers of $\beta$. On the basis of simpler cosmological models, one can argue that the effective Hamiltonian can be obtained from the classical one by substituting even powers of $v\beta$ with the same powers of the function
\begin{subequations}\label{eq:lqc-polymer}
\begin{equation}\label{eq:omega}
\Omega=v\frac{\sin(\sqrt{\Delta}\beta) }{\sqrt{\Delta}},
\end{equation}
where $\Delta=4\sqrt{3}\pi\gamma\ell^2_{\rm Pl}$ is the minimum eigenvalue allowed for the area operator in loop quantum gravity (here, $\gamma$ is the Immirzi parameter, and $\ell^2_{\rm Pl}=G\hbar$ is the square of the Planck length). As for the odd powers of the form $(v \beta)^{2k+1}$, which were treated differently in the quantization prescriptions of Refs.~\cite{fmo1,fmo2}, we have chosen to replace them with $\Omega^{2k}\Lambda$, where
\begin{equation}\label{eq:lambda}
\Lambda = v\frac{\sin(2\sqrt{\Delta}\beta) }{2\sqrt{\Delta}}.
\end{equation}
\end{subequations}

After introducing the variables~\eqref{eq:lqc-variables} and the polymerization scheme of Eqs.~\eqref{eq:lqc-polymer}, the reduced system is subject to the Hamiltonian constraint
\begin{equation}\label{eq:total-const}
H(\bar N_0)=\frac{\bar  N_0}{16\pi G}\Big(C_0+\sum_{\vec n,\epsilon} C_2^{\vec n,\epsilon}\Big),
\end{equation}
with the lapse $\bar N_0=\sigma N_0$, the unperturbed Hamiltonian constraint
\begin{equation}\label{eq:C0-const}
C_0=-\frac{6}{\gamma^2}\frac{\Omega^2}{v}
+\frac{ 8\pi G}{v}\pi_\phi^2,
\end{equation}
and the quadratic contribution of each of the modes
\begin{equation}\label{eq:C2-const}
C_2^{\vec n,\epsilon}=\frac{8\pi G}{v^{1/3}}\left(\tilde E^n_{\pi\pi}\tilde \pi^2_{\tilde f_{\vec n,\epsilon}}+2\tilde E^n_{f\pi}\tilde f_{\vec n,\epsilon}\tilde \pi_{\tilde f_{\vec n,\epsilon}}+\tilde E^n_{ff}\tilde f_{\vec n,\epsilon}^2\right).
\end{equation}
The coefficients in the above equation are given by
\begin{subequations}\label{eq:Es}
\begin{align}
\tilde E^n_{\pi\pi} &= 1-\frac{4\pi G}{\tilde\omega_n^2}\frac{\pi_\phi^2}{v^{4/3}},\\
\tilde E^n_{f\pi} &= -\frac{8\pi G}{\tilde \omega_n^2}\frac{\pi_\phi^2\Lambda}{\gamma v^2},\\\nonumber
\tilde E^n_{ff} &= \tilde \omega_n^2-\frac{1}{2v^{4/3}}\left[\frac{\Omega^2}{\gamma^2}+20\pi G\pi_\phi^2 \right]\\
&-\frac{16\pi G}{\tilde \omega_n^2v^{8/3}}\left(\frac{\pi_\phi\Lambda}{\gamma}\right)^2,
\end{align}
\end{subequations}
with $\tilde \omega_n^2=4\pi^2\vec n\cdot\vec n$. Another convenient description of the contribution of the perturbations to the global Hamiltonian constraint is given in terms of the creation and annihilation variables that are naturally associated to the massless representation adopted in Refs.~\cite{fmo,fmo1,fmo2}, namely,
\begin{eqnarray}\label{eq:creat-like}
a_{\tilde f_{\vec n,\epsilon}}=\frac{1}{\sqrt{2\tilde \omega_n}}(\tilde \omega_n  \tilde f_{\vec n,\epsilon}+i\tilde \pi_{\tilde f_{\vec n,\epsilon}})
\end{eqnarray}
and their complex conjugates $a^*_{\tilde f_{\vec n,\epsilon}}$, where now
\begin{equation}\label{eq:inh_quan_const}
C^{\vec n,\epsilon}_2=H_0^{\vec n,\epsilon}+H_{\mathrm{int}}^{\vec n,\epsilon}
\end{equation}
with
\begin{align}
&H_0^{\vec n,\epsilon}=\frac{8\pi G}{v^{1/3}}N^{\vec n,\epsilon}\bigg(2\tilde \omega_n+\frac{1}{\tilde \omega_n} F^n_{-}\bigg),\\
&H_{\mathrm{int}}^{\vec n,\epsilon} =\frac{4\pi G}{\tilde\omega_nv^{1/3}}\bigg( X_+^{\vec n,\epsilon}  F^n_{+}+i\frac{4\pi G}{\tilde \omega_n} X_-^{\vec n,\epsilon}  G^n\bigg).
\end{align}
Here,
\begin{eqnarray}\label{eq:CoeffA}
N^{\vec n,\epsilon}&=& a^*_{\tilde f_{\vec n,\epsilon}}  a_{\tilde f_{\vec n,\epsilon}}, \quad  X_\pm^{\vec n,\epsilon}= ( a^*_{\tilde f_{\vec n,\epsilon}})^2\pm  a_{\tilde f_{\vec n,\epsilon}}^2,\\ \nonumber
F_\pm^n&=&-
\frac{1}{2v^{4/3}}\left[\frac{\Omega^2}{\gamma^2}+4\pi G(5\mp 2) \pi_\phi^2\right]
\\
&&-\frac{16\pi G}{\tilde \omega_n^2v^{8/3}}\bigg(\frac{\pi_\phi\Lambda}{\gamma }
\bigg)^2,\\
G^n&=&-\frac{4}{\gamma v^2} \pi^2_\phi\Lambda .
\end{eqnarray}

\subsection{Equations of motion}

In the Hamilton equations of motion, for the sake of consistency with our truncation scheme, we will consider
the backreaction on the zero modes owing to quadratic contributions of the perturbations in the action. Hence,
\begin{subequations}\label{eq:hom-eqs}
\begin{align}
&\dot{\phi} = \bar N_0\frac{\pi_\phi}{v}
\nonumber\\ &+\frac{\bar N_0}{v^{1/3}} \sum_{\vec n,\epsilon} \left(\tilde E^n_{\phi,\pi\pi}\tilde \pi^2_{\tilde f_{\vec n,\epsilon}}+2\tilde E^n_{\phi,f\pi}\tilde f_{\vec n,\epsilon}\tilde \pi_{\tilde f_{\vec n,\epsilon}}+\tilde E^n_{\phi,ff}\tilde f_{\vec n,\epsilon}^2\right),\\\label{eq:dot-piphi}
&\dot{\pi}_\phi = 0,
\\\label{eq:dot-vol}
&\dot v = \frac{3}{2}\bar N_0v\frac{\sin(2\sqrt{\Delta}\beta)}{\sqrt{\Delta}\gamma}\nonumber\\
&+\frac{\bar N_0}{v^{1/3}}\sum_{\vec n,\epsilon} \left(2\tilde E^n_{v,f\pi}\tilde f_{\vec n,\epsilon}\tilde \pi_{\tilde f_{\vec n,\epsilon}}+\tilde E^n_{v,ff}\tilde f_{\vec n,\epsilon}^2\right),\\\nonumber
& \dot \beta = -\frac{3}{2}\bar N_0\frac{\sin^2(\sqrt{\Delta}\beta)}{\Delta\gamma}-2\pi G\gamma \bar N_0 \frac{\pi_\phi^2}{v^2}- \frac{\gamma}{12}\frac{\bar N_0}{v}\sum_{\vec n,\epsilon}  C_2^{\vec n,\epsilon} \\
&+\frac{\bar N_0}{v^{1/3}}\sum_{\vec n,\epsilon} \left(\tilde E^n_{\beta,\pi\pi}\tilde \pi^2_{\tilde f_{\vec n,\epsilon}}+2\tilde E^n_{\beta,f\pi}\tilde f_{\vec n,\epsilon}\tilde \pi_{\tilde f_{\vec n,\epsilon}}+\tilde E^n_{\beta,ff}\tilde f_{\vec n,\epsilon}^2\right),
\end{align}
\end{subequations}
where the functions $\tilde E^n$ are obtained by taking the Dirac brackets of the homogeneous variables with the coefficients~\eqref{eq:Es}. Explicitly,
\begin{subequations}
\begin{align}
\tilde E^n_{\phi,\pi\pi} &= -\frac{8\pi G\pi_\phi}{\tilde \omega_n^2v^{4/3}}, \\
\tilde E^n_{\phi,f\pi} &= -\frac{16\pi G\pi_\phi\Lambda}{\tilde \omega_n^2v^2\gamma }, \\
\tilde E^n_{\phi,ff} &= -\frac{4\pi G}{v^{4/3}}\left[5\pi_\phi+\frac{8\pi_\phi \Lambda^2}{\tilde \omega_n^2v^{4/3}\gamma^2}\right], \\
\label{eq:E_vfp}
\tilde E^n_{v,f\pi} &= \frac{32\pi^2 G^2\pi_\phi^2}{\tilde\omega_n^2v}\cos(2\sqrt\Delta\beta), \\
\label{eq:E_vff}
\tilde E^n_{v,ff} &= \frac{4\pi G\Lambda}{v^{1/3}\gamma }\left[\frac{32\pi G\pi_\phi^2}{\tilde \omega_n^2 v^{4/3}}\cos(2\sqrt{\Delta}\beta)+1\right], \\
\tilde E^n_{\beta,\pi\pi} &= \frac{64\pi^2 G^2\gamma\pi_\phi^2}{3\tilde \omega_n^2 v^{7/3}}, \\
\tilde E^n_{\beta,f\pi} &= \frac{32\pi^2 G^2\pi_\phi^2\Lambda}{\tilde \omega_n^2v^3}, \\
\tilde E^n_{\beta,ff} &= \frac{4\pi G \gamma}{3v^{7/3}}\left[40\pi G\pi_\phi^2-\frac{\Omega^2}{\gamma^2}
+\frac{32\pi G \pi_\phi^2\Lambda^2}{\tilde \omega_n^2v^{4/3}\gamma^2}\right].
\end{align}
\end{subequations}

The time evolution of the perturbations, on the other hand, is dictated by an infinite number of first order differential equations:
\begin{align}\label{eq:lon-g-pert}\nonumber
\dot{\tilde f}_{\vec n,\epsilon} &=\frac{\bar N_0}{v^{1/3}}(\tilde E^n_{\pi\pi}\tilde \pi_{\tilde f_{\vec n,\epsilon}}+\tilde E^n_{f\pi}\tilde f_{\vec n,\epsilon}),\\\dot{\tilde \pi}_{\tilde f_{\vec n,\epsilon}} &=-\frac{\bar N_0}{v^{1/3}}(\tilde E^n_{ff}\tilde f_{\vec n,\epsilon}+\tilde E^n_{f\pi}\tilde \pi_{\tilde f_{\vec n,\epsilon}}),
\end{align}
which do not mix different modes and (given our perturbative truncation) are linear in the inhomogeneities.

In addition to the previous equations, the variables must satisfy the constraint equation,
\begin{equation}\label{eq:const-equat}
C_0+\sum_{\vec n,\epsilon} C_2^{\vec n,\epsilon}=0,
\end{equation}
where $C_0$ and $C_2^{\vec n,\epsilon}$ are given in Eqs.~\eqref{eq:C0-const} and \eqref{eq:C2-const}, respectively.

We see from Eq.~\eqref{eq:dot-piphi} that, in this particular cosmological model (with a massless scalar field) and up to the perturbative order considered here, the momentum $\pi_\phi$ is a constant of motion, just as in the unperturbed model.

\subsubsection{Bounce and critical energy density}\label{sec:bounce-rhoc}

In the system that we are considering, the energy density varies locally on each spatial section because of inhomogeneity, but we can still define an averaged effective energy density which includes not only the contribution of the homogeneous matter field, but also that of the metric and matter perturbations averaged on each section:
\begin{subequations}
\begin{align}
\rho_\mathrm{eff} &= \rho+\tilde\rho,\\
\rho&=\frac{\pi_{\phi}^2}{2v^2}, \quad \tilde{\rho}:=\frac{1}{16\pi Gv}\sum_nC_2^{\vec n,\epsilon}.\label{eq:energy-den}
\end{align}
\end{subequations}
According to the constraint equation~\eqref{eq:const-equat}, $\rho_\mathrm{eff}$ reaches its maximum $\rho_c= 3/(8\pi\gamma^2 G\Delta)$ when $\sqrt{\Delta}\beta=\pi/2$. If the backreaction of the perturbations on the homogeneous variables is negligible or can be disregarded, the homogeneous part of the volume reaches its maximum at the same point of the evolution, since $\dot{v}$ vanishes when $\sqrt{\Delta}\beta=\pi/2$, in agreement with Eq.~\eqref{eq:dot-vol}. However, if one incorporates the backreaction as given by Eqs.~\eqref{eq:E_vfp} and \eqref{eq:E_vff}, the two events no longer coincide. This fact is confirmed by the numerical simulations (see Sec.~\ref{sec:numerics}) and can be traced back to the specific prescription that we have adopted for substituting the odd powers of $v\beta$ in the construction of the effective dynamics of the model.

The minimum volume $v_b$ can be calculated by setting $\dot v$ [Eq.~\eqref{eq:dot-vol}] equal to zero and solving for $v$. The solution is determined by an equation of the form
\begin{equation}
v_b^{7/3}+av_b^{5/3}+bv_b^{1/3}+c=0,
\end{equation}
where the coefficients $a$, $b$, and $c$ depend on the values of $\beta$, $\pi_\phi$, and the inhomogeneities at the bounce. In the simulations that we have performed, given two trajectories with and without inhomogeneities but with the same value of the momentum $\pi_\phi$, the volume at the bounce $v_b$ is slightly lower in our inhomogeneous system than in the unperturbed scenario. More details about the numerical results can be found in Sec.~\ref{sec:numerics}.

\subsection{The inhomogeneous sector and the massless representation}

In this section we consider two alternate descriptions of the inhomogeneous sector. The first one is provided by the scaled modes of the matter field, introduced above. In the ultraviolet limit ($\omega_n\rightarrow\infty$), the dynamics of the variables $\tilde f_{\vec n,\epsilon}$ approximates that of the modes of a Klein-Gordon field subject to a quadratic potential that depends on time through the homogeneous variables.

The other description that we consider below is given in terms of gauge invariants of general relativity. Far away from the quantum regime, where the effective dynamics is well approximated by general relativity, the configuration variable of this alternate description satisfies also equations of the Klein-Gordon form with a time dependent quadratic potential.

We show below that the limit of the aforementioned potentials in the asymptotic past and future is flat, and hence the field behaves asymptotically as a massless scalar field on a static Minkowski spacetime in these descriptions, in agreement with the numerical results shown in Figs.~\ref{fig:figtimepot} and~\ref{fig:figenergpert}.

\subsubsection{Scaled matter perturbation}\label{sec:scaledpert}

Taking into account Eq.~\eqref{eq:lon-g-pert}, one can easily arrive at the second order differential equation (see Ref.~\cite{fmo}):
\begin{align}\label{eq:KG-like}
0 =& \tilde f''_{\vec n,\epsilon}-\frac{\tilde E^{n\prime}_{\pi\pi}}{\tilde E^n_{\pi\pi}}\tilde f'_{\vec n,\epsilon} \\
\nonumber
&+\left(\frac{\tilde E^{n\prime}_{\pi\pi}}{\tilde E^n_{\pi\pi}} \tilde E^n_{f\pi}- \tilde E^{n\prime}_{f\pi}-\big( \tilde E^n_{f\pi}\big)^2+ \tilde E^n_{ff} \tilde E^n_{\pi\pi}\right) \tilde f_{\vec n,\epsilon},
\end{align}
where the primes indicate derivatives with respect to the conformal time, defined by $d\eta=\bar N_0v^{-1/3}dt$.

The term multiplying $\tilde f_{\vec n,\epsilon}'$ in this dynamical equation can be checked to be negligible in the ultraviolet limit ($\tilde\omega_n\rightarrow\infty$). Actually, this fact motivated the choice of the variables~\eqref{eqs:newvariablesB}. As for the term multiplying  $\tilde f_{\vec n,\epsilon}$, it can be decomposed as the sum of the contribution of the Laplacian, $\tilde\omega_n^2$, and a function $V_{\tilde f_{\vec n,\epsilon}}$, which depends on time and also in the frequency of the mode. In the numerical simulations that we have performed, this term is well defined almost everywhere in the evolution. In fact, the existence of some points where this quantity is ill behaved is not problematic for our analysis, since the equations that we actually integrate are Eqs.~\eqref{eq:lon-g-pert}, which are free of any divergence. In the next section, we show how our numerical results confirm that the function $V_{\tilde f_{\vec n,\epsilon}}$ tends to zero in both the asymptotic past and future. Although the numerical study is particularized there to the lowest mode, the conclusion is general. In fact, the ultraviolet limit of $V_{\tilde f_{\vec n,\epsilon}}$ admits a simple expression (see App.~\ref{ap:potentials} for additional details). Thus, the high frequency scaled modes of the field behave approximately like those of a scalar field with a quadratic potential whose coefficient is given by (half) that limit. Neglecting holonomy corrections ($\Lambda\simeq v\beta\simeq \Omega$), and using the constraint at the perturbative order that is relevant here, one can check that the behavior of this potential in asymptotic regions of large volume $v$ is determined by
\begin{equation}\label{eq:timedeppot}
\lim_{\tilde{\omega}_n\rightarrow\infty}V_{\tilde f_{\vec n,\epsilon}} \simeq -\frac{44\pi G\pi_\phi^2}{3v^{4/3}},
\end{equation}
so it tends to zero as $v^{-4/3}$. Consequently, the differential equations satisfied by the modes $f_{\vec n,\epsilon}$ reduce in the asymptotic limit to those corresponding to a massless field on a static, flat spacetime.

\subsubsection{Mukhanov-Sasaki variable}

The gauge invariant Mukhanov-Sasaki variable plays a prominent role in cosmology, not only because of its simple relation with the perturbed scalar curvature on the spatial hypersurfaces, but also because, in general relativity, it behaves as a Klein-Gordon field on a static spacetime interacting with a time dependent potential in flat, isotropic cosmologies, where it satisfies the equation
\begin{equation}\label{eq:Mukhanov-Sasaki}
v_{\vec n,\epsilon}'' + \left(\tilde\omega_n^2-\frac{z''}{z}\right)v_{\vec n,\epsilon}= 0.
\end{equation}
Here, the function $z$ can be defined as
\begin{equation}\label{eq:ms-z}
z = \frac{\gamma\pi_\phi}{v^{2/3}\beta}.
\end{equation}
The expression of this gauge invariant and its canonically conjugate momentum in terms of the scaled perturbation of the matter field were given in Ref.~\cite{fmo2} and they can also be found in App.~\ref{ap:gauge-inv}. In the effective system, Eq.~\eqref{eq:Mukhanov-Sasaki} holds only if holonomy corrections are negligible. Within that regime, the time dependent potential that rules the evolution of the Mukhanov-Sasaki variable is determined by the function
\begin{equation}\label{eq:tdp-MS}
-\frac{z''}{z} \simeq -\frac{4\pi G\pi_\phi^2}{3v^{4/3}}.
\end{equation}
In order to arrive at the above expression, we have used the Hamiltonian constraint (see App.~\ref{ap:potentials}). Since the right hand side vanishes in the limit $v\to \infty$, the modes of the gauge invariant potential $v_{\vec n,\epsilon}$ behave asymptotically as those of a massless Klein-Gordon field on a static, flat spacetime.

\section{Numerical results}\label{sec:numerics}

We have performed the numerical integration of solutions that correspond to an initially collapsing Universe which experiences a bounce in the evolution and expands from there on. In doing this, we have employed an adaptive Runge-Kutta method 89 (see Ref.~\cite{verner}). Besides, in all simulations we have chosen units so that the values of the Newton and the Plank constants are $G=1/(8\pi)$ and $\hbar = 1$, respectively, and we have fixed the Immirzi parameter equal to $\gamma = 0{.}23753295797$ (keeping just the digits that are  
significant at the level of the numerical error of our simulations). Our results are summarized below.

\subsection{Evolution of the perturbations neglecting backreaction contributions}

\begin{figure}
\includegraphics[scale=0.75]{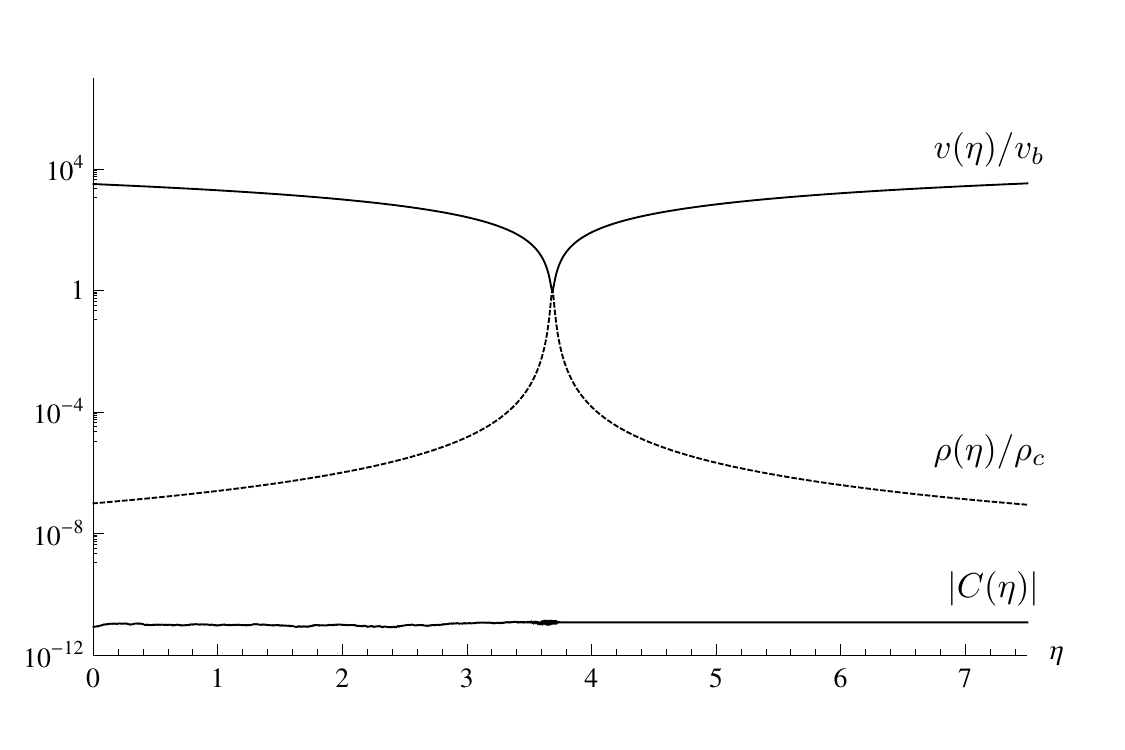}
\caption{\scriptsize We plot, in conformal time, the normalized volume $v(\eta)/v_b$ (where $v_b$ is the value of the volume at the bounce), the normalized energy density $\rho(\eta)/\rho_c$ (where $\rho_c$ is the critical energy density), and the absolute value of the constraint $|C(\eta)|$. Our initial data are $v=10^5$, $\phi=0{.}1$, and
$\beta=0{.}9999\pi/\sqrt{\Delta}$, whereas $\pi_\phi$ is determined by using the constraint.
}    \label{fig:figvpphi}
\end{figure}

\begin{figure*}
\centering \subfigure[]{\label{fig:figtimepota}
\includegraphics[scale=0.75]{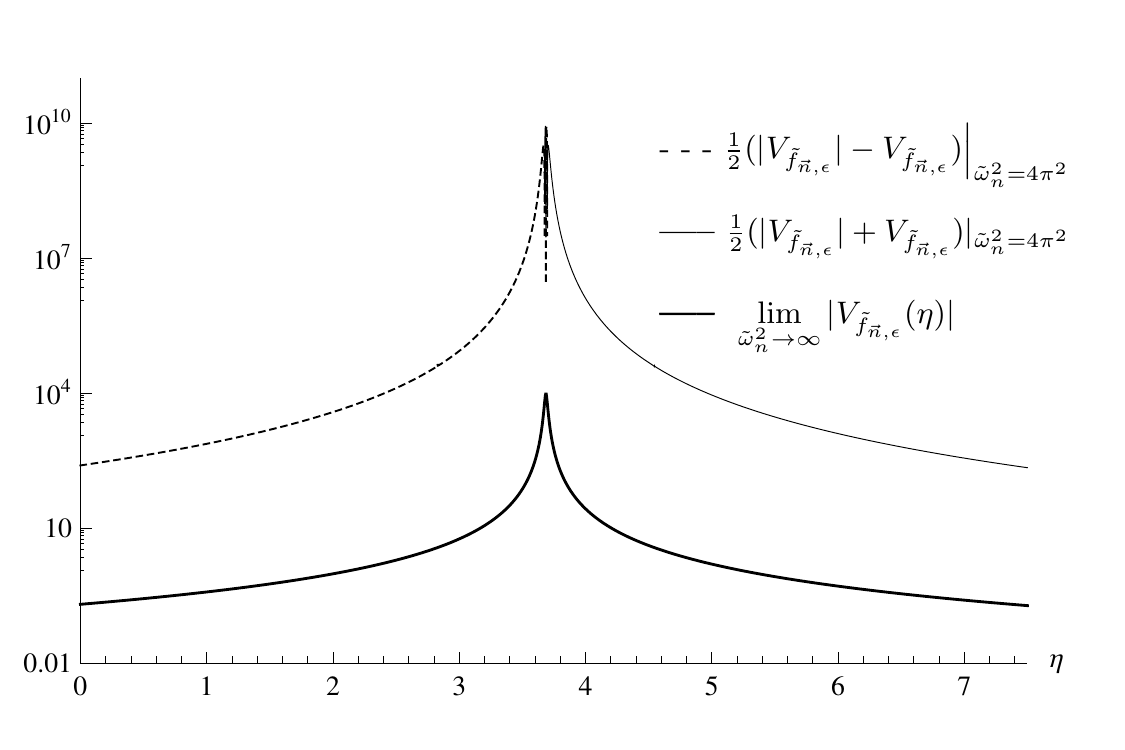} } \subfigure[]{\label{fig:figtimepotb}
\includegraphics[scale=0.75]{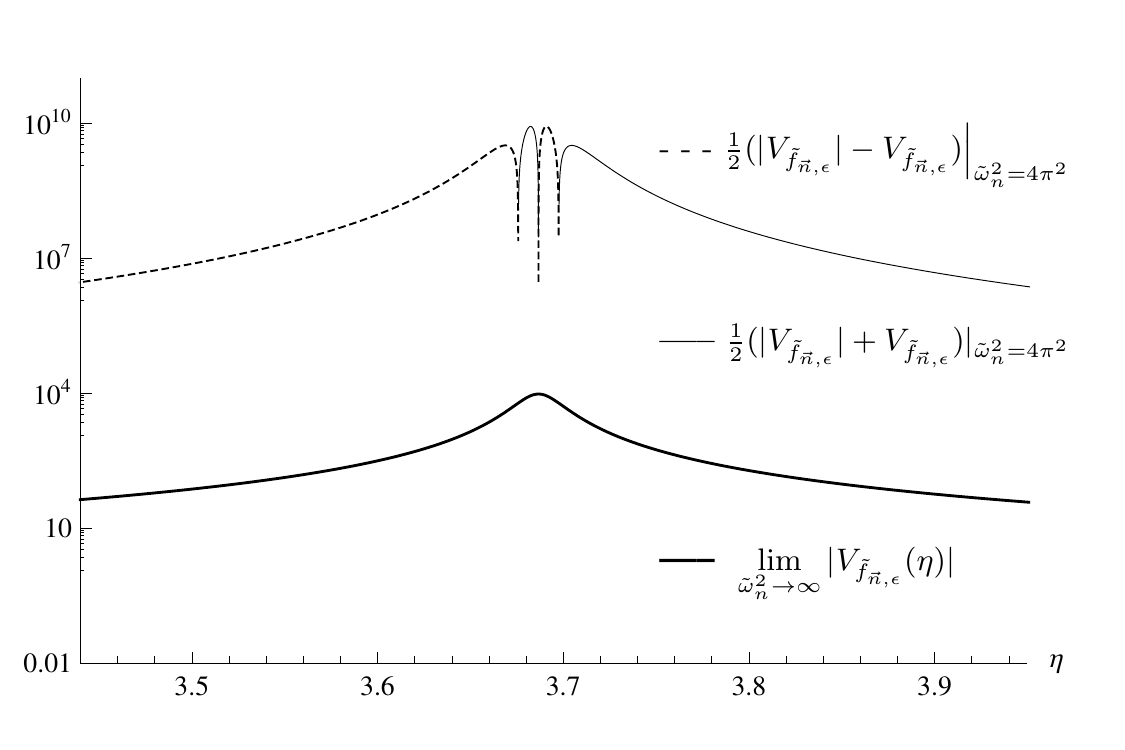} } \caption{\scriptsize (a) Time dependent
coefficient $V_{\tilde{f}_{\vec n,\epsilon}}(\eta)$ for the ultraviolet modes and the
lowest mode $\tilde\omega^2_n=4\pi^2$. (b) Zoom of the plot around the bounce
region, in conformal time $\eta$. The initial
data are $v=10^5$, $\phi=0{.}1$, and $\beta=0{.}9999\pi/\sqrt{\Delta}$, whereas $\pi_\phi$
is determined by means of the constraint.
}   \label{fig:figtimepot}
\end{figure*}

Let us start with a particular solution in which all backreaction contributions are small enough to be disregarded. Figure~\ref{fig:figvpphi} shows the evolution of the homogeneous physical volume $v$, which allows us to identify a bounce, together with the variation of the energy density, defined in Eq.~\eqref{eq:energy-den}. In addition, we include the actual value of the scalar constraint, given by the sum of Eq.~\eqref{eq:C0-const} and of
Eq.~\eqref{eq:C2-const} for each mode. It vanishes up to numerical errors.

In Fig.~\ref{fig:figtimepot}, we plot the ultraviolet limit of the function $V_{\tilde f_{\vec n,\epsilon}}$ (defined in Sec.~\ref{sec:scaledpert}), which can be interpreted as the coefficient of a time dependent quadratic potential that rules the evolution of the ultraviolet modes of the matter perturbation. This limit is given by Eq.~\eqref{eq:timedeppot}. We also plot the value of $V_{\tilde f_{\vec n,\epsilon}}$ for the lowest mode, namely $\tilde\omega_n^2=4\pi^2$. We have not displayed the evolution of the corresponding function for the modes of the gauge invariant Mukhanov-Sasaki variable [see Eq.~\eqref{eq:Mukhanov-Sasaki}] because it behaves similarly to $V_{\tilde{f}_{\vec n,\epsilon}}$ in the limit $v\to\infty$, up to constant factors. Notice the logarithmic scale employed for the ordinate axis.

\begin{figure*}
\centering
\subfigure[]{\label{fig:figenergpert1}
\includegraphics[scale=0.75]{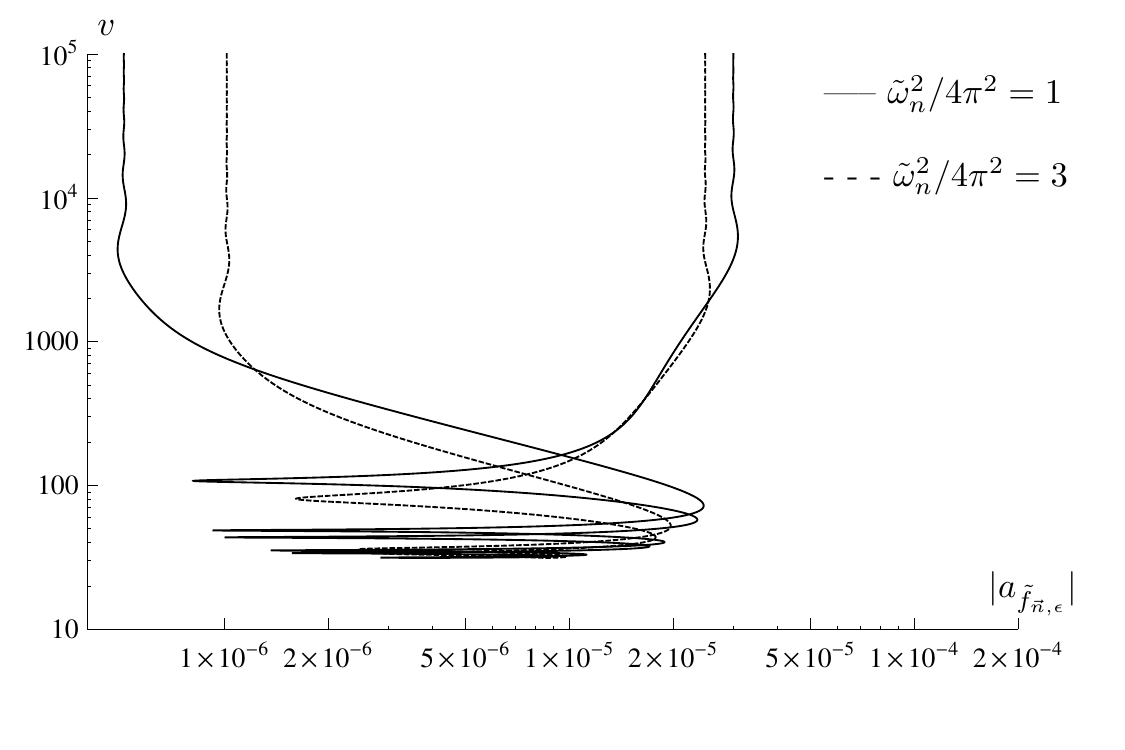}
}
\subfigure[]{\label{fig:figenergpert2}
\includegraphics[scale=0.75]{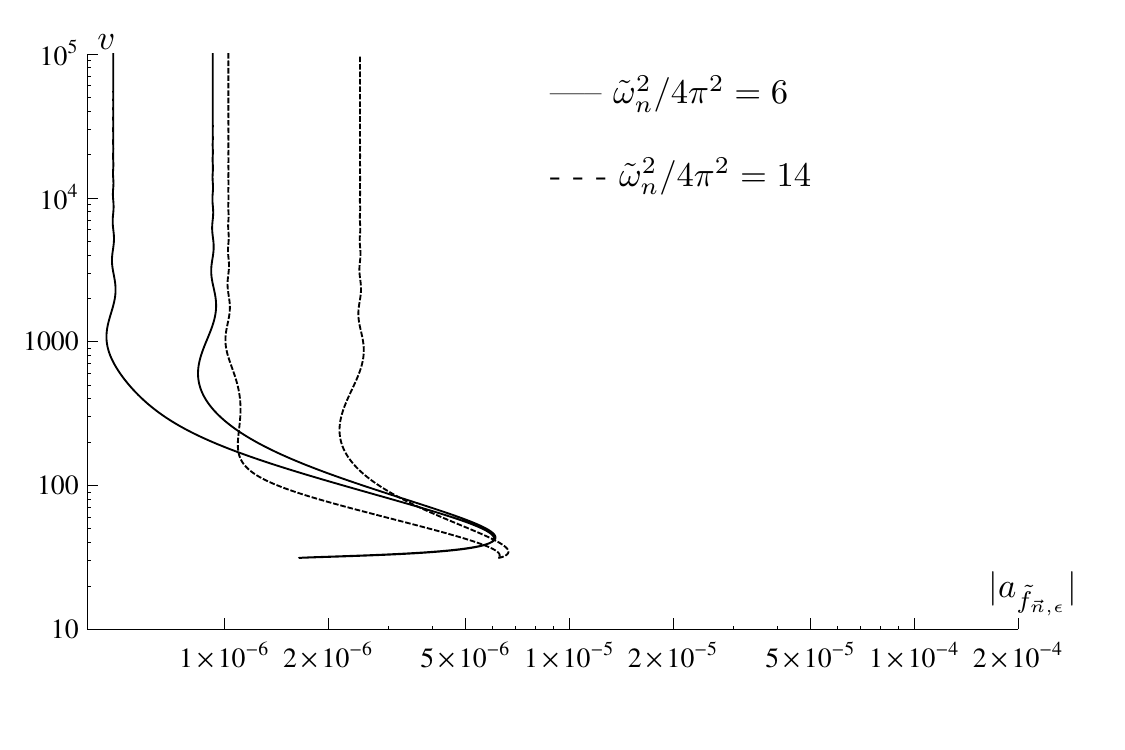}
}
\caption{\scriptsize Evolution of the perturbations for modes whose amplitude is
considerably amplified after the bounce (a) and for modes that are just
slightly amplified (b). Given the (massless) annihilation variable $a_{\tilde{f}_{\vec
n,\epsilon}}$ in the longitudinal gauge, we plot $|a_{\tilde{f}_{\vec n,\epsilon}}|$ in
the abscissa with respect to the volume $v$. Our initial data are $v=10^5$, $\phi=0{.}1$, and $\beta=0{.}9999\pi/\sqrt{\Delta}$, whereas
$\pi_\phi$ is determined by means of the constraint. The initial data for
the perturbations are $a_{\tilde{f}_{\vec n,\epsilon}} = \epsilon e^{i\alpha}$ and
$a^*_{\tilde{f}_{\vec n,\epsilon}} = \epsilon e^{-i\alpha}$, with $\epsilon$ and
$\alpha$ randomly distributed, following respectively a normal distribution of
zero mean and dispersion equal to $10^{-6}$, and a flat distribution in $(0,2\pi]$.}
\label{fig:figenergpert}
\end{figure*}

Note also that these functions can become negative on the considered trajectory. We see that they tend to zero at the asymptotic past and future; however, close to the bounce they are in general non-negligible (and negative). Therefore, one can expect that the oscillatory behavior of the perturbations will give way to an amplification regime at a certain stage around the bounce, especially for the infrared (low) modes. This is in fact corroborated in Fig.~\ref{fig:figenergpert}, where we plot the absolute value of $a_{\tilde{f}_{\vec n,\epsilon}}$ for several modes of the perturbation. We can observe two types of behavior. In Fig.~\ref{fig:figenergpert1} the quantity $|a_{\tilde{f}_{\vec n,\epsilon}}|$, for each mode, oscillates around a mean value in both the distant past of the collapsing branch of the Universe (left) and the distant future of the expanding branch (right). The mean values in the latter have increased by more than one order of magnitude when compared to those in the former branch. On the other hand, in Fig.~\ref{fig:figenergpert2}, we observe that there are some modes whose amplitude is only slightly amplified.

In order to understand the characteristics of this amplification mechanism in more detail, we have studied it from a statistical point of view \emph{neglecting all backreaction effects}. We follow here a strategy similar to that presented in Ref.~\cite{bmp}. Specifically, we start with a particular background trajectory and we parameterize the inhomogeneities in the following way:
\begin{equation}
a_{\tilde{f}_{\vec n,\epsilon}}(\eta)=|a_{\tilde{f}_{\vec n,\epsilon}}(\eta)|e^{i\alpha(\eta)},
\end{equation}
where the variables $a_{\tilde{f}_{\vec n,\epsilon}}$ are defined in Eq.~\eqref{eq:creat-like} for each mode. They are the annihilation variables associated to a massless vacuum state (see Ref.~\cite{fmo}). We will assume that initially the term proportional to $\tilde f_{\vec n,\epsilon}$ in Eq.~\eqref{eq:KG-like} is dominated by the contribution of the Laplacian, $\tilde\omega^2_n$, and that the perturbations are initially in the adopted massless vacuum state. We then consider samples of initial data for these perturbations such that the amplitudes $|a_{\tilde{f}_{\vec n,\epsilon}}(\eta_0)|$ ($\eta_0$ being an arbitrary initial time) follow a normal distribution of zero mean and dispersion equal to $\sigma$, while the phases $\alpha(\eta_0)$ are selected according to a uniform distribution on the interval $(0,2\pi]$. Therefore, the parameter $\sigma$ governs the initial amplitude of the perturbations. We have run simulations with two types of initial conditions, either i)~considering one mode for each of the chosen values of $\tilde{\omega}_n$ and averaging over different simulations, or ii)~considering the degeneracy of the eigenspaces corresponding to each of the chosen $\tilde{\omega}_n$'s and averaging over all possible modes as well as over different simulations. Our study shows that both approaches give the same qualitative results. Regarding the statistical analysis, each mode is always averaged over samples with more than one hundred elements.

\begin{figure*}[tbh!]
\subfigure[]{
\includegraphics[scale=0.76]{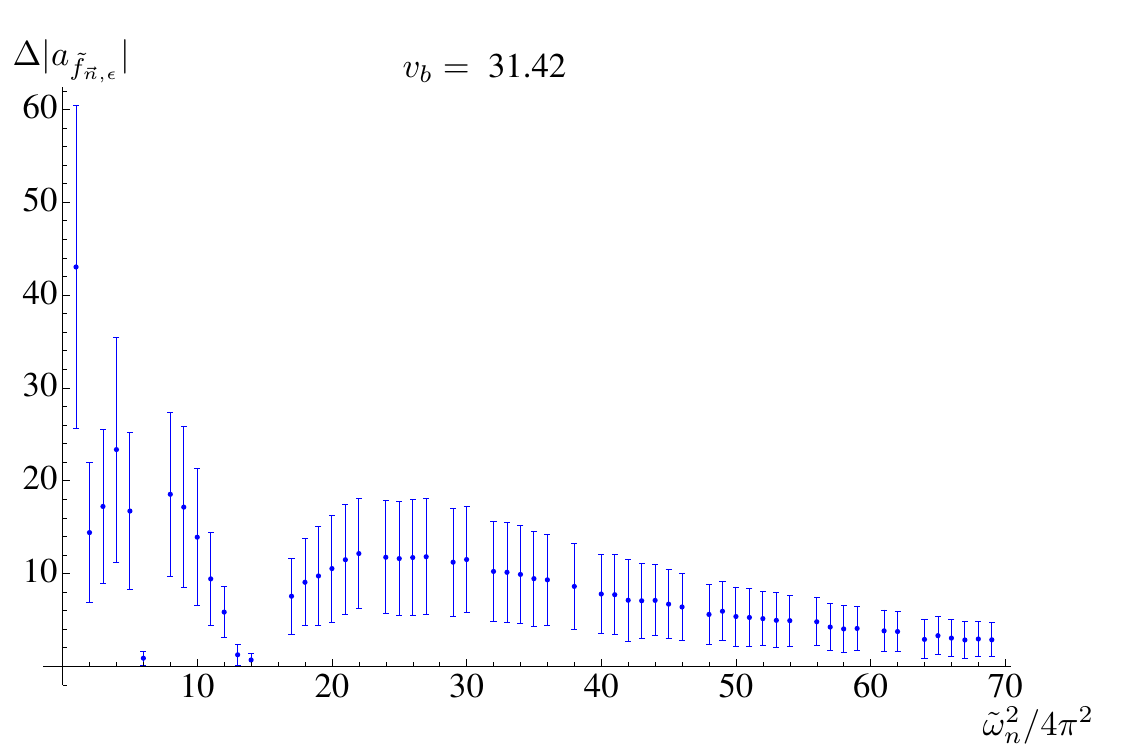}
}
\subfigure[]{
\includegraphics[scale=0.76]{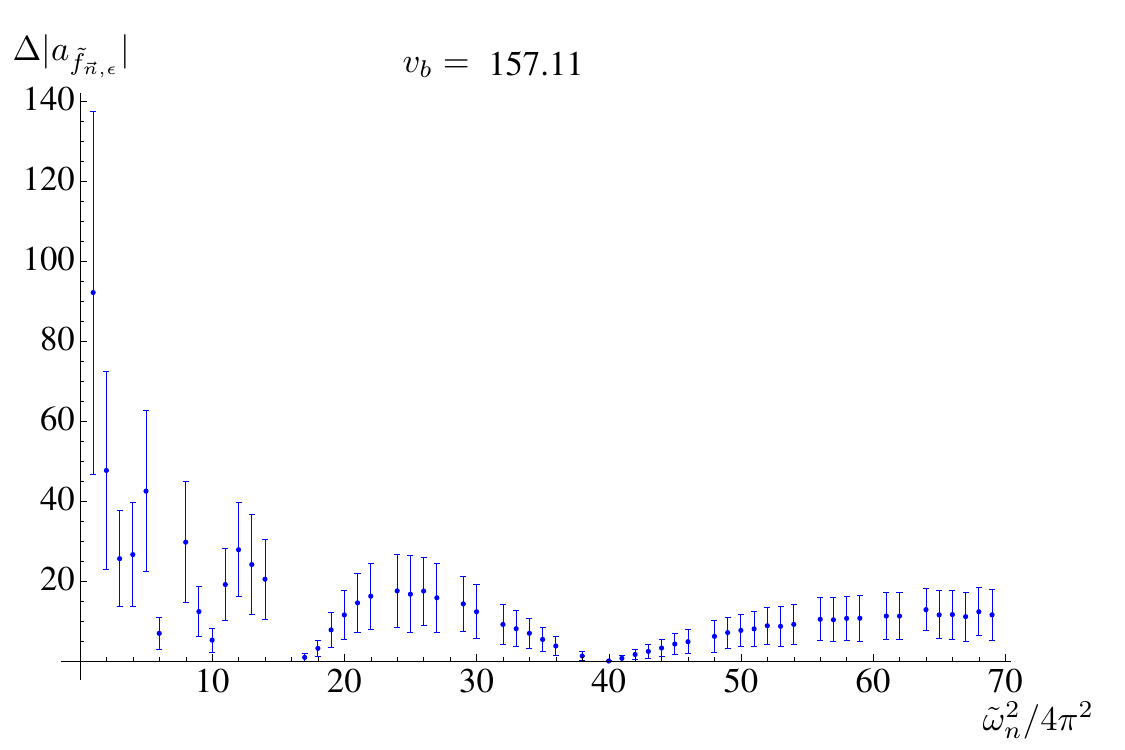}
}
\subfigure[]{
\includegraphics[scale=0.8]{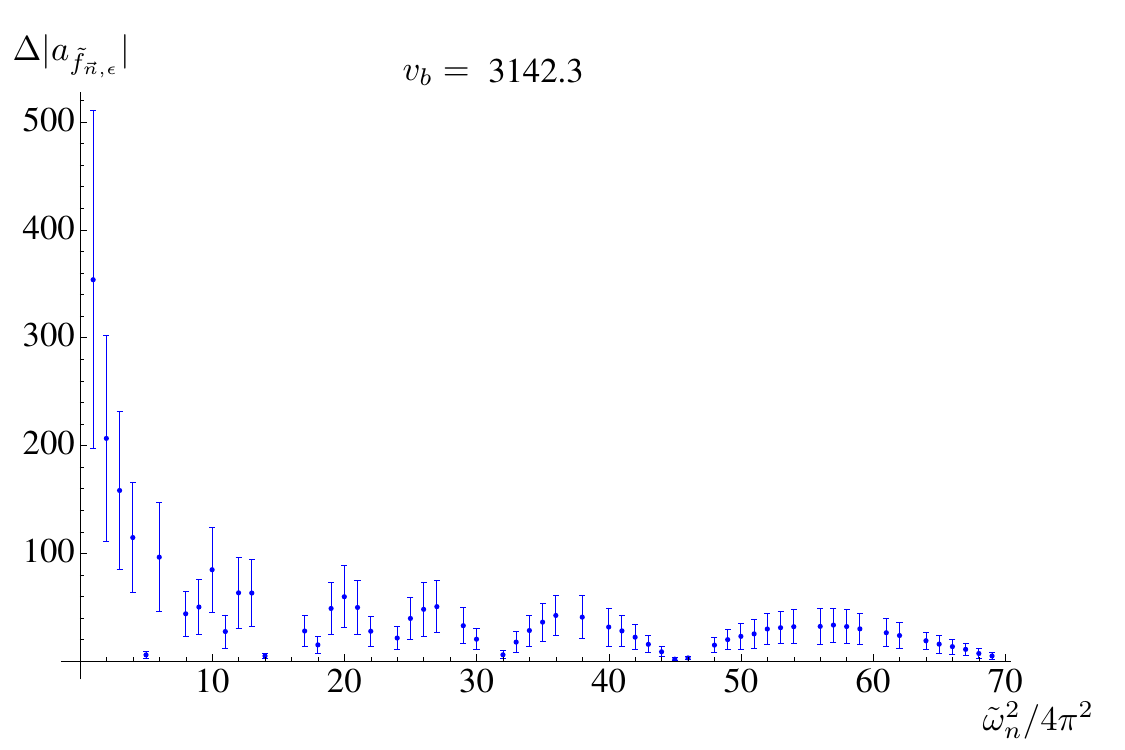}
}
\caption{\scriptsize Statistical amplification of the modes after the bounce. For these simulations, we take $\phi=0{.}1$ and different values of $v$ and $\beta$  at the initial time
($\pi_\phi$ is determined by means of the constraint). The initial data for the
perturbations are $a_{\tilde{f}_{\vec n,\epsilon}} = \epsilon e^{i\alpha}$  and
$a^*_{\tilde{f}_{\vec n,\epsilon}} = \epsilon e^{-i\alpha}$, with $\epsilon$ and $\alpha$
randomly distributed (following respectively a normal distribution of zero mean
and dispersion $10^{-10}$, and a flat distribution in $(0,2\pi]$). The bars show the standard deviation over modes with the same eigenvalue.}
\label{fig:statistics}
\end{figure*}

The main results are illustrated in Fig.~\ref{fig:statistics}, where we have plotted the mean value and dispersion of the observable
\begin{equation}
\Delta |a_{\tilde{f}_{\vec n,\epsilon}}|:=\frac{|a_{\tilde{f}_{\vec n,\epsilon}}(\eta_\mathrm{f})|-|a_{\tilde{f}_{\vec n,\epsilon}}(\eta_0)|}{|a_{\tilde{f}_{\vec n,\epsilon}}(\eta_0)|}
\end{equation}
with respect to the frequency $\tilde{\omega}_n$ (or, more exactly, to its square). Here, $\eta_\mathrm{f}$ and $\eta_0$ are
certain values of the conformal time after and before the bounce,
respectively (and sufficiently far from it so as to eliminate transient regimes in the amplitude variation). The perturbations, on average, are amplified, and the amplification depends on the particular homogeneous trajectory under consideration and on the concrete mode eigenspace. Besides, we have checked that the amplification statistics are insensible to the initial amplitudes $|a_{\tilde{f}_{\vec n,\epsilon}}(\eta_0)|$ of the mode.
Of course, this fact is due to the linearity of the equations of motion of the perturbations and the negligibility of the backreaction. However, it is worth noticing that the amplification depends on the initial phases $\alpha(\eta_0)$, which have nonetheless been sampled with flat distributions in our statistical analysis. As for the amplification factor, we see that it is bigger for trajectories with bigger values of the volume $v$ at the bounce, and it decays to the unit (and hence the amplitude variation to zero) for ultraviolet modes. This behavior is in agreement with that of the function $V_{\tilde{f}_{\vec n,\epsilon}}$, which decays to zero asymptotically but takes non-negligible negative values around the bounce which affect importantly the evolution of the infrared modes.

\begin{figure*}[tbh!]
\subfigure[]{
\includegraphics[scale=0.76]{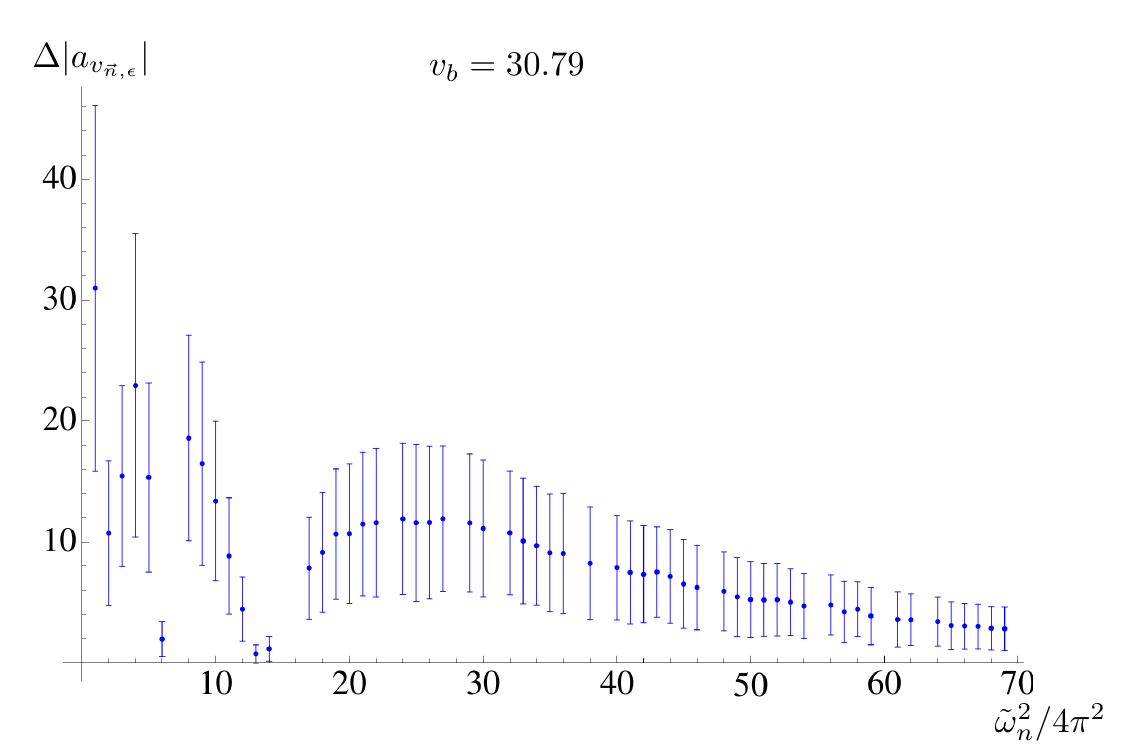}
}
\subfigure[]{
\includegraphics[scale=0.76]{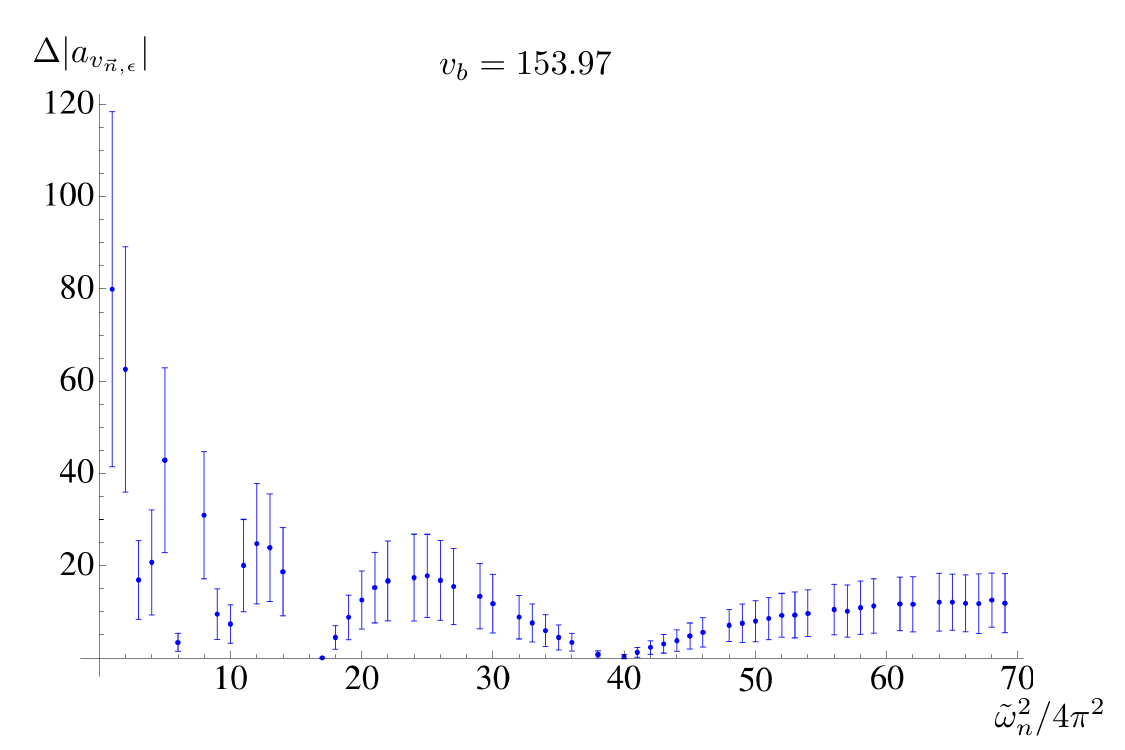}
}
\subfigure[]{
\includegraphics[scale=0.8]{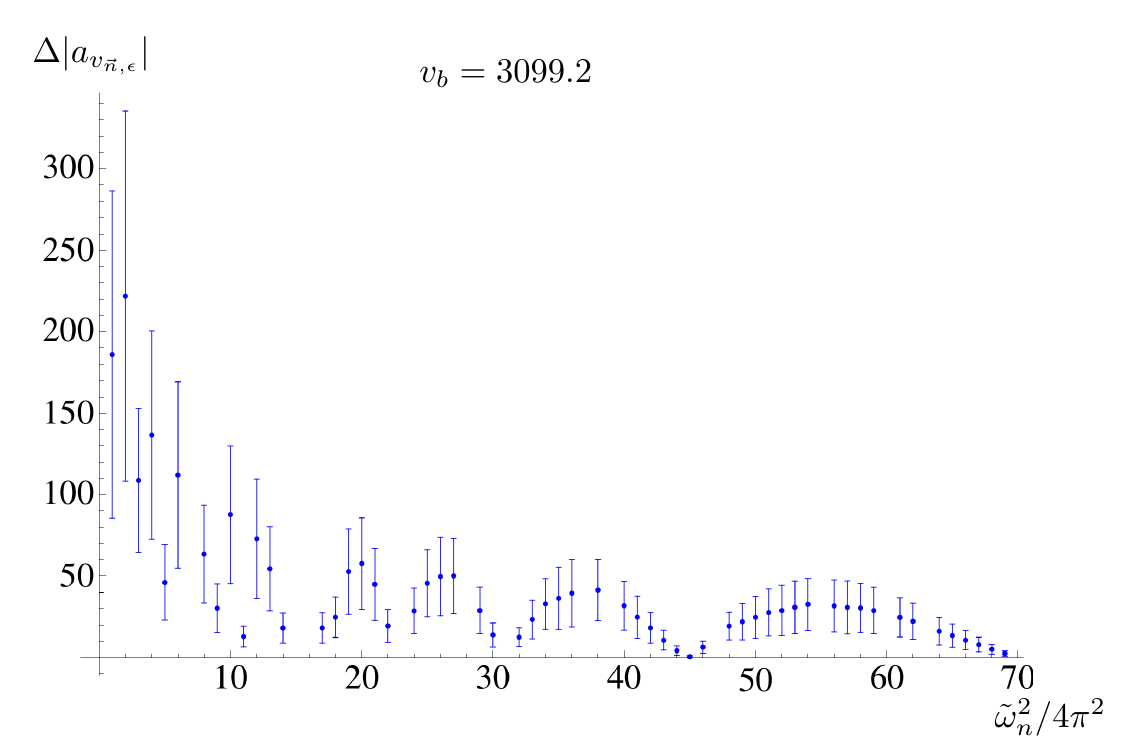}
}
\caption{\scriptsize Statistical amplification of the modes of the gauge invariant Mukhanov-Sasaki variable after the bounce. We
take $\phi=0{.}1$ and different values of $v$ and $\beta$ as initial conditions for the homogeneous sector
(with $\pi_\phi$ determined by the constraint). For the
perturbations, we consider initial data of the form $a_{v_{\vec n,\epsilon}} = \epsilon e^{i\alpha}$  and
$a^*_{v_{\vec n,\epsilon}} = \epsilon e^{-i\alpha}$, with $\epsilon$ and $\alpha$
randomly distributed (following respectively a normal distribution of zero mean
and dispersion $10^{-10}$, and a flat distribution in $(0,2\pi]$).}
\label{fig:statistics-vn}
\end{figure*}

Moreover, in Fig.~\ref{fig:statistics} we can see a superposed oscillatory pattern with a period that seems to behave qualitatively as $\log\tilde\omega_n$, and which also depends on the specific background trajectory under consideration. These oscillations were unexpected for us, based just on the present limited experience with this kind of systems. In order to check that they are not an artifact of the gauge fixing, we have carried out a similar analysis for the gauge invariant Mukhanov-Sasaki variable. The results are summarized in Fig.~\ref{fig:statistics-vn}, where we observe qualitatively the same amplification pattern and oscillations. We plan to study whether they persist in the genuine inflationary scenario (with massive scalar fields), since they might give rise to testable predictions of quantum gravity phenomena.


\begin{figure*}[tbh!]
\subfigure[]{\includegraphics[width=0.49\linewidth]{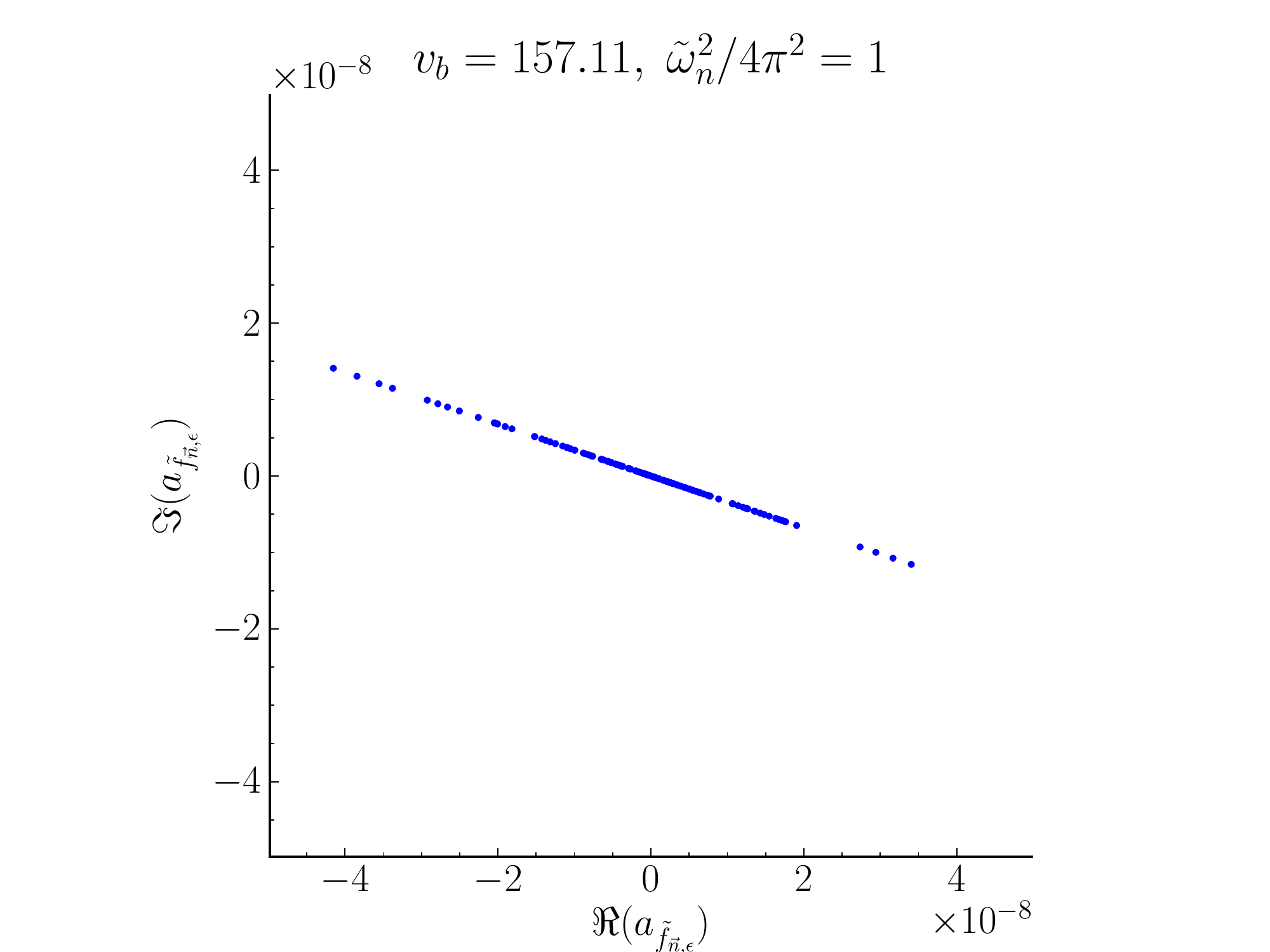}\label{fig:phasesa}}
\subfigure[]{\includegraphics[width=0.49\linewidth]{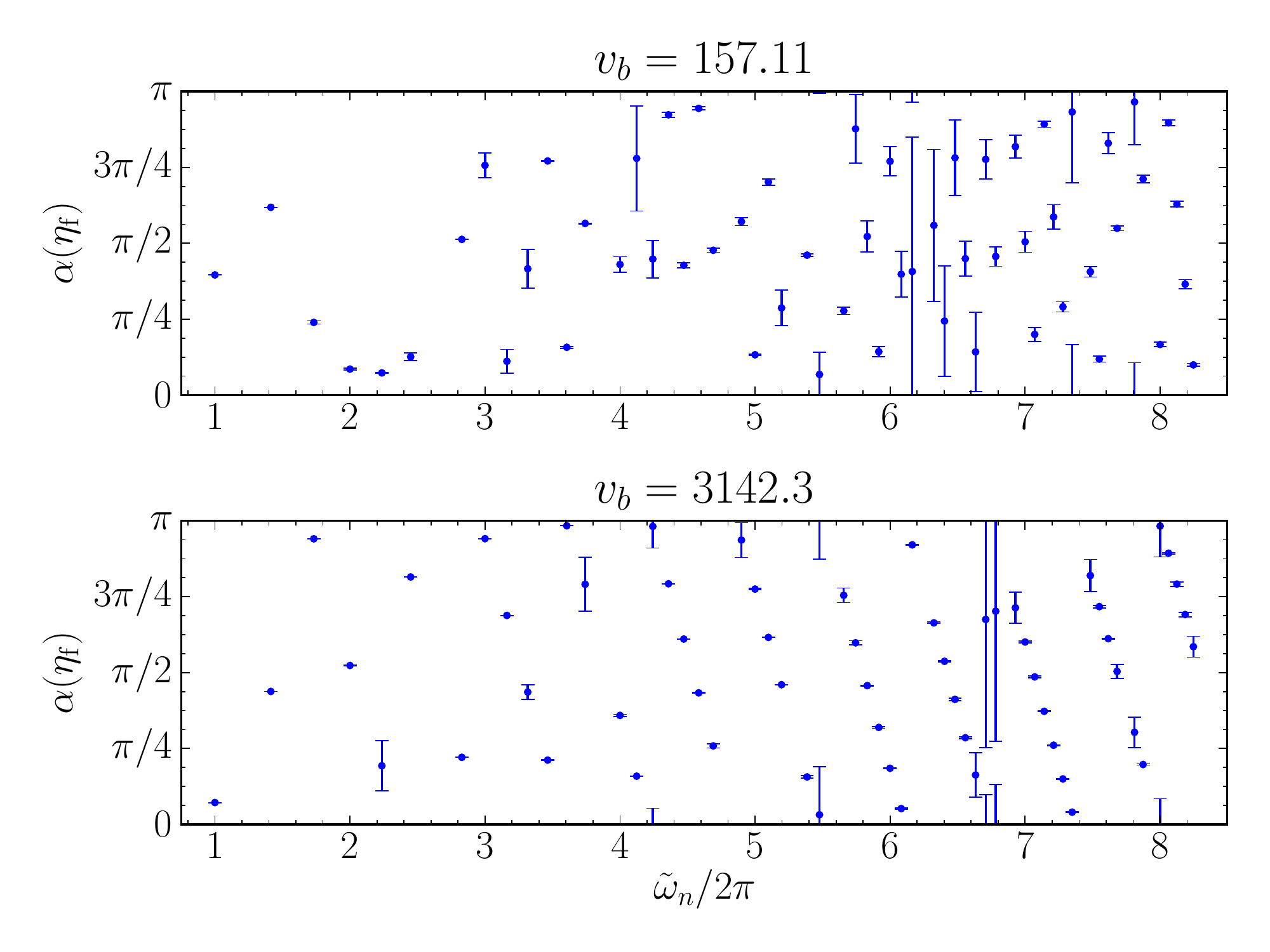}\label{fig:phasesb}}
\caption{\scriptsize Alignment of the phases of the perturbations after the bounce. Initial data have been chosen as for Fig.~\ref{fig:statistics}. The final time $\eta_\mathrm{f}$ was selected so that $v(\eta_0)=v(\eta_\mathrm{f})$ in each case. (a) The value of the variable $a_{\tilde{f}_{\vec n,\epsilon}}(\eta_\mathrm{f})$ in different simulations has been represented in the complex plane for a typical mode with small dispersion of the phases. Each point corresponds to the same mode in different simulations. (b) Statistical analysis of the phases $\alpha_n(\eta_\mathrm{f})$. By convention, $\alpha_n\in(-\pi,\pi]$. The bars show the standard deviation corresponding to each eigenvalue. }
\label{fig:phases}
\end{figure*}

On the other hand, since time reversal is a symmetry of the effective system under consideration, the statistical amplification of the perturbations must have a counterpart in the distribution of their phases. The simulations have confirmed that this is indeed the case. Given a final time $\eta_\mathrm{f}$ well after the bounce, the distribution of the final phases $\alpha_n(\eta_{\mathrm{f}})$ is peaked around two opposite angles (differing in $\pi$ radians). The exact value of these angles of accumulation depends on the considered mode in a simple way (it is approximately linear in $\tilde\omega_n$), although the dependence pattern is obscured by the periodicity of the angles. The dispersion around the angles of accumulation is also mode dependent. This is illustrated in Fig.~\ref{fig:phases}. The modes with higher dispersions are those with lower amplification factor, as one might have expected. It would be interesting to explore possible consequences of these effects in the phase distribution of the perturbations that might be falsifiable by comparison with observational data. This will be the subject of further research.

\subsection{Evolution with backreaction}

\begin{figure*}
\centering
\subfigure[]{
\includegraphics[scale=0.76]{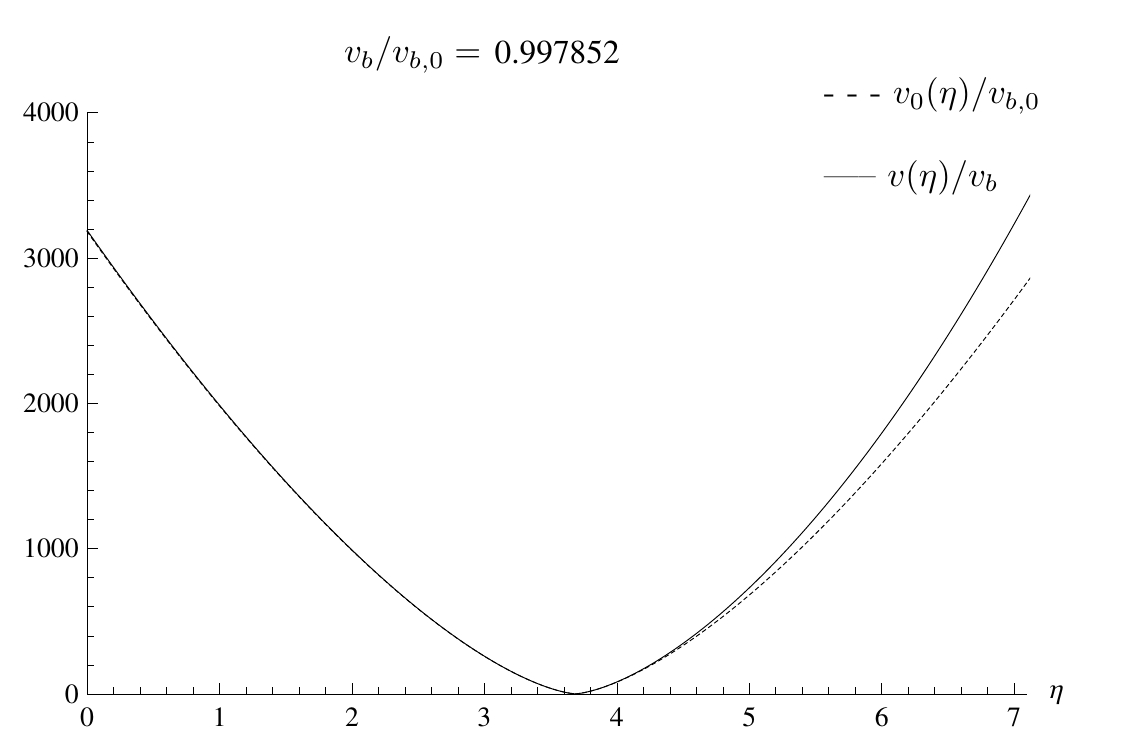}     \label{fig:BR1a}
}
\subfigure[]{
\includegraphics[scale=0.76]{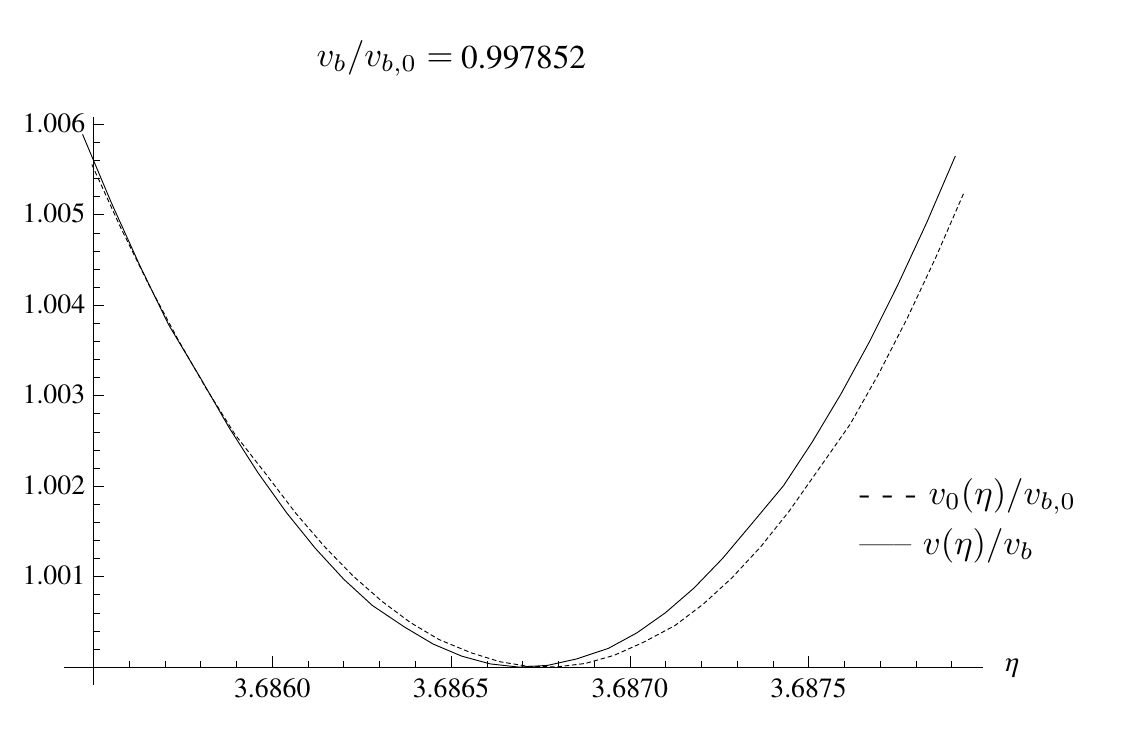}     \label{fig:BR1b}
}
\caption{\scriptsize (a) Normalized {\it homogeneous} physical volume of the Universe with and without perturbations ($v/v_{b}$ and $v_0/v_{b,0}$, respectively). (b) Zoom at the bounce. In these simulations we take the initial conditions $\phi=0{.}1$, $v=10^5$, and $\beta=0{.}9999\pi/\sqrt{\Delta}$ ($\pi_\phi$ is determined by means of the constraint). Therefore, $v_{b,0}=31{.}4159$. The initial data for the perturbations are $a_{\tilde{f}_{\vec n,\epsilon}} = \epsilon e^{i\alpha}$ and $a^*_{\tilde{f}_{\vec n,\epsilon}} = \epsilon e^{-i\alpha}$, with $\epsilon$ and $\alpha$ randomly distributed (following respectively a normal distribution of zero mean and dispersion $10^{-2}$, and a flat distribution in $(0,2\pi]$). All modes with $\tilde\omega^2\in[1,6]$ have been considered, including degeneration. }    \label{fig:BR1}
\end{figure*}

In addition to the numerical analysis presented in the previous section, we have run simulations that include the backreaction of the proposed effective dynamics. We describe here the main qualitative results by considering two families of trajectories, with and without inhomogeneities, which share the same initial data for $v$, $\phi$, and $\pi_\phi$ [in this case $\beta$ is fixed by the constraint \eqref{eq:total-const}]. The initial amplitudes of the perturbations were selected randomly, following a normal distribution of zero mean, as before, but with a higher dispersion, equal to $10^{-2}$. Their initial phases were chosen again with a flat distribution in $(0,2\pi]$. This choice in fact has no physical relevance for the analysis. Figure~\ref{fig:BR1} displays the evolution of the homogeneous physical volume of the Universe, $v$, normalized by the corresponding value at its bounce, in presence and absence of perturbations. The origins of time have been chosen so that both trajectories reach the critical energy density at the same time. Figure~\ref{fig:BR1a} shows that the inhomogeneities accelerate the growth of the physical volume $v$ once the bounce is over.

Besides, we observe in Fig.~\ref{fig:BR1b} that the value of the homogeneous volume $v$ at its bounce changes in the presence of inhomogeneities ---although this change depends on the particular distribution of those inhomogeneities--- and that there is a time delay. These effects are due to backreaction. Nevertheless, let us comment that the observed delay is a direct consequence of the quantization prescription adopted in Ref.~\cite{fmo2} and the related assumption for the replacement of odd powers of $v\beta$ in the effective dynamics [Eq.~\eqref{eq:lambda}], and therefore it would not be present in other prescriptions.

\begin{figure}
\centering
\includegraphics[scale=0.76]{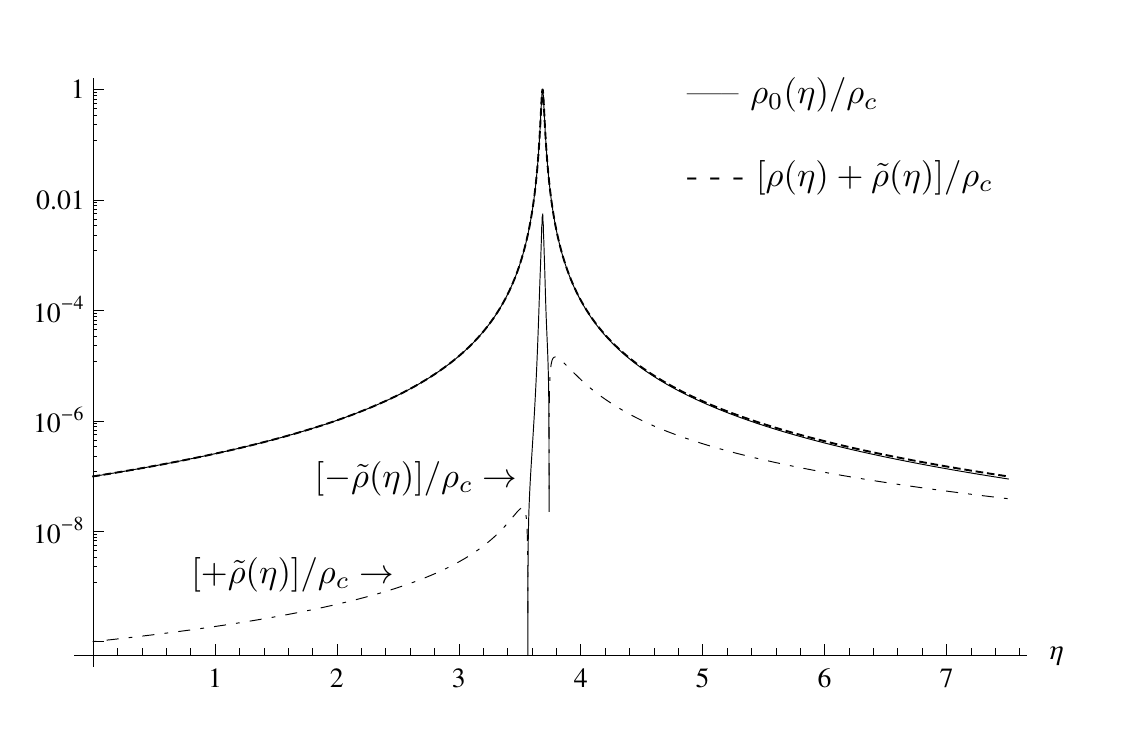}
\caption{\scriptsize Energy density of the background $\rho_0/\rho_c$ without perturbations, total effective energy density $(\rho+\tilde\rho)/\rho_c$, and absolute value of the energy density $|\tilde\rho|/\rho_c$ corresponding to the perturbations, all of them normalized with respect to the critical energy density $\rho_c$. Initial data have been chosen as for Fig.~\ref{fig:BR1}. All modes with $\tilde\omega^2\in[1,6]$ have been taken into account, including degeneration.}    \label{fig:BR2}
\end{figure}

In Fig.~\ref{fig:BR2} we display the energy density $\rho$ of the zero mode of the matter field when the inhomogeneities are absent, the total effective energy density  $\rho+\tilde\rho$, and the one corresponding only to the perturbations $\tilde\rho$ [see Eq.~\eqref{eq:energy-den}], all of them normalized by the critical energy density $\rho_c$. The value of $\tilde\rho$ is negative close to the bounce, whereas it becomes comparable to $\rho$ in the expanding period, a fact which proves that the criterion suggested in Ref.~\cite{AAN2} in order to neglect the backreaction contribution is not fulfilled in this particular case, even if $\tilde\rho$ is nonetheless small enough to permit disregarding higher order contributions. We see that the total effective energy density for both solutions is qualitatively similar.

\begin{figure*}
\centering
\subfigure[]{
\includegraphics[scale=0.76]{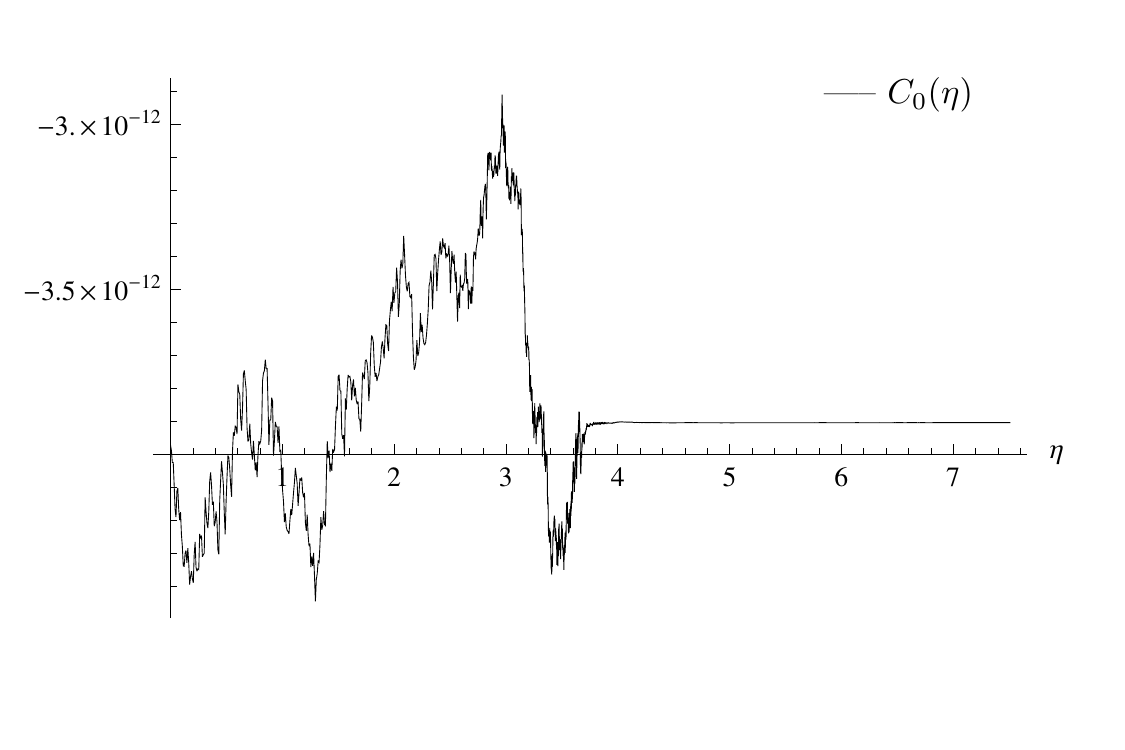}
}
\subfigure[]{
\includegraphics[scale=0.76]{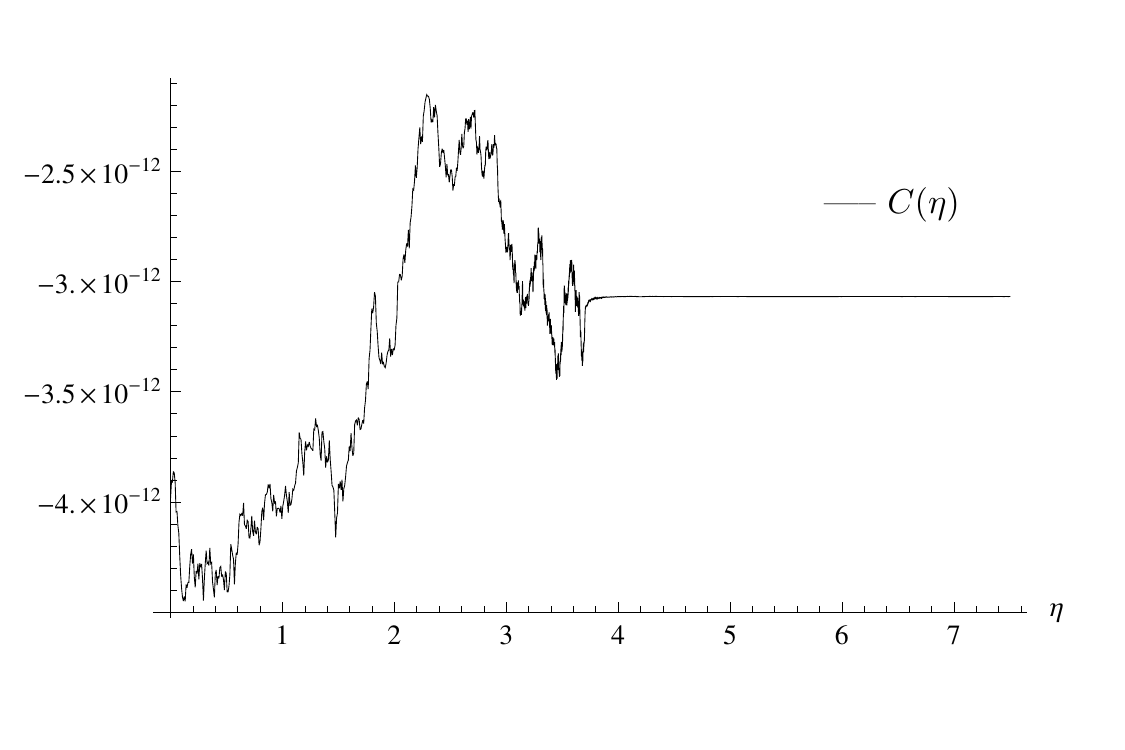}
}
\caption{\scriptsize (a) Numerical value of the constraint $C_0(\eta)$ for an unperturbed solution; (b) Numerical value of the complete constraint $C(\eta)$, with initial data for the perturbations given by $a_{\tilde{f}_{\vec n,\epsilon}} = \epsilon e^{i\alpha}$ and $a^*_{\tilde{f}_{\vec n,\epsilon}} = \epsilon e^{-i\alpha}$, where $\epsilon$ and $\alpha$ are randomly distributed (following respectively a normal
distribution of zero mean and dispersion $10^{-2}$, and a flat distribution in $(0,2\pi]$). All modes with $\tilde \omega^2\in[1,6]$ have been considered, including degeneration. The initial values adopted in these simulations for the homogeneous variables are $\phi=0{.}1$, $v=10^5$, and $\beta=0{.}9999\pi/\sqrt{\Delta}$ ($\pi_\phi$ is determined by means of the
constraint).}   \label{fig:BR3}
\end{figure*}

For the sake of completeness, we also include in Fig.~\ref{fig:BR3} the numerical evaluation of the constraint for both cases (homogeneous and inhomogeneous scenarios). One can see that the corresponding constraints are satisfied up to numerical errors.

\section{Conclusions}\label{sec:conclu}

In this work we have studied a flat FRW spacetime coupled to a scalar field and small inhomogeneities introduced as perturbations, treated up to quadratic order in the action. The inhomogeneous model possesses a global scalar constraint, and two local constraints that are linear in the perturbations. The first one is composed of two contributions: one depending exclusively on the zero modes of the metric and matter fields and the other containing the perturbations. We have adopted the longitudinal gauge and introduced new phase space variables that are canonical for the reduced model. The reduced system is subject only to a global scalar constraint. Our choice of variables for it is motivated by the quantum description proposed in Refs.~\cite{fmo,fmo1,fmo2}, which combines a standard Fock representation for the matter content together with a loop quantization for the geometry. Then, we have analyzed a proposal for the effective dynamics of the quantum model based on the results of loop quantum cosmology in simpler systems \cite{effecti}, straightforwardly generalized to inhomogeneous scenarios like the one considered here. We have included in the description the backreaction of the inhomogeneities on the background variables up to the order of our perturbative truncation. We have studied the possible influence of the backreaction in our effective model, finding that it alters the growth of the homogeneous physical volume $v$ in the expanding branch of the Universe. In addition, the minimum of $v$ is modified. On the other hand, this minimum and the maximum of the effective energy density $\rho_\mathrm{eff}$ do not occur simultaneously in our simulations. However, this effect may be a consequence of the proposal adopted for the effective dynamics, and hence its analysis must not be considered conclusive.

Besides, we have carried out a statistical study in order to understand how the perturbations are affected by the quantum corrections encoded in the effective dynamics when the backreaction can be safely neglected. In particular, we have evolved random initial conditions before the bounce, as if the inhomogeneities were initially in a vacuum state of the massless representation~\cite{fmo}, and we have compared these data with those resulting from their evolution at the other side of the Universe, namely in the expanding branch. We have observed that, in most of the cases, the infrared (i.e., the low) modes of the perturbations suffer an amplification of several orders of magnitude in their amplitude. This phenomenon can be understood qualitatively by thinking of the perturbation as a field which evolves approximately as a Klein-Gordon field with a negative time dependent potential that influences significantly the dynamics in the region of the bounce. This amplification does not affect all modes in the same way, and in fact it fades out in the ultraviolet limit. Moreover, the amplification factor presents oscillations in its dependence on the mode frequency. It would be interesting to explore if these oscillations leave a trace that might be falsified with observational data; this will be a matter for future analysis.

The amplification of the inhomogeneities through the bounce was already found in the hybrid quantization of the Gowdy model~\cite{bmp} and also appears in other treatments of cosmological perturbations within loop quantum cosmology (for instance, the case of tensor perturbations is investigated in Ref.~\cite{LCBG13}). Furthermore, in fact this feature has appeared in other bouncing cosmologies (see, e.g., Ref.~\cite{bounce}), unrelated to a loop formulation of gravity. A novelty here is the modulation of the amplification factor with the frequency, whose possible observational implications will be studied elsewhere.

On the other hand, the amplification of the perturbations seems to indicate an energy transfer from the background to the inhomogeneities. We have found that this is indeed the case in trajectories with appreciable (but sufficiently small) backreaction: the corresponding energy is provided by the zero mode of the matter field, whose energy density becomes lower after the bounce, owing precisely to backreaction effects. The total effective energy density (zero mode and inhomogeneities) is approximately similar to that found in the absence of perturbations.

Another interesting phenomenon concerns the effect of the amplification mechanism on the phases of the different modes. We have observed that certain modes with the same dynamics tend to emerge from the quantum regime with the very same phase, regardless of their initial phases. In this strict sense, the information encoded before the bounce in those modes phases (considered apart from the rest of the system) is wiped out in the bouncing process. Nonetheless, there are other modes that are not affected by the amplification nor by the alignment of phases. In any case, this phenomenon deserves a careful analysis in more realistic scenarios, particularly in genuine inflationary settings, to elucidate whether it actually persists and its possible physical consequences.

Let us comment that our choice of the longitudinal gauge in this work has been intentional, since this gauge fixing provides a reduced system whose Hamilton equations are well defined in the whole reduced phase space (except at the singular point $v=0$). In this way, we have avoided dangerous numerical instabilities in the solution space (where the singular point $v=0$ is never reached). This is not the case, e.g., for the other alternate gauge fixing considered in Refs.~\cite{fmo,fmo1,fmo2}, in which the singular behavior on the hypersurface $\pi_{\alpha}=0$ introduces considerable difficulties in order to attain a reliable numerical integration. Moreover, we have repeated the numerical analysis of the amplification of the perturbations but adopting instead a description in terms of the gauge invariant Mukhanov-Sasaki variable (see App.~\ref{ap:gauge-inv}), checking in this way that the observed effect is not an artifact of the gauge fixing. The same type of amplification patterns have been reproduced with this gauge invariant.

We can be sure of the robustness of our numerical estimations for several reasons. We have integrated the Hamilton equations \eqref{eq:hom-eqs} and \eqref{eq:lon-g-pert} employing both the Runge-Kutta 45 and 89 algorithms. In both cases we have considered different choices for the tolerance controlling the adaptative stepsize, obtaining the same qualitative results. Besides, during the integration, we have evaluated the scalar constraint at all time steps and we have used its absolute value as a numerical tolerance function. More specifically, we have fixed its maximum admissible value to be smaller than the tolerance parameter $\delta$, with $\delta\in[10^{-10},10^{-7}]$ depending on the numerical trajectory.

Once we have analyzed the evolution of the perturbations in a simple background scenario (filled with a massless scalar field), we are ready to carry out the same analysis for the inflationary scenario in which the scalar field is massive. A preliminary numerical analysis has shown that, during the slow-roll regime, a straightforward integration of the equations of motion becomes unstable. In absence of backreaction, however, the system can be in principle integrated within the slow-roll approximation. Therefore, it seems possible to carry out a statistical analysis of the amplification of the perturbations {similar to that presented here but now on inflationary trajectories, with an eye in searching for observationally relevant} consequences. Nonetheless, if the backreaction is sufficiently intense during the slow-roll period, additional considerations (and/or reformulations of the system) would be necessary in order to deal satisfactorily with the numerical integration of the dynamics. This will be a subject of future research.

\section*{Acknowledgements}

The authors are grateful to D. Mart\'{\i}n-de Blas for discussions and for his help with numerical integrations. They are also greatly thankful to T. Paw{\l}owski for conversations and for allowing the use of numerical codes employed in the simulations. This work was supported by PEDECIBA (Uruguay) and the Projects No.\ MICINN/MINECO FIS2011-30145-C03-02 and CPAN CSD2007-00042 from Spain. M. F.-M. acknowledges CSIC and the European Social Fund for support under the Grant No.\ JAEPre\_2010\_01544.

\appendix

\section{Harmonics}\label{ap:harm}

In order to deal with the spatial dependence of the perturbations, it is extremely convenient to expand them in terms of the eigenbases of (scalar, vector, and tensor) harmonics of the Laplace-Beltrami operator corresponding to the reference three-metric. In the case under consideration, we introduce the \emph{real} eigenfunctions
\begin{equation} \label{Qharmo}
\tilde Q_{\vec n,+} = \sqrt 2\cos\left(\frac{2\pi}{l_0}\vec n\cdot\vec\theta\right),\quad  \tilde Q_{\vec n,-} = \sqrt 2\sin\left(\frac{2\pi}{l_0}\vec n\cdot\vec\theta\right),
\end{equation}
where $\vec n$ is any tuple of integers whose first
non-vanishing component is positive (we exclude non-zero modes from the perturbations). Notice that this basis of scalar modes is
normalized so that
\begin{equation}
\int_{T^3}\!\!d^3\theta\,\tilde Q_{\vec n,\epsilon}(\vec\theta)\tilde Q_{\vec n',\epsilon'}(\vec\theta) = l_0^3\,\delta_{\vec n,\vec n'}\delta_{\epsilon,\epsilon'},
\end{equation}
with $\epsilon,\epsilon'=+,-$. The corresponding eigenvalue equation for the Laplace-Beltrami operator is
\begin{equation}
^0h^{ij}(\tilde Q_{\vec n,\epsilon})_{|ij} = -\omega_n^2\tilde Q_{\vec n,\epsilon}.
\end{equation}

We can construct vector and tensor modes from these scalar ones by covariant differentiation. In this work, we do not include genuine vector and tensor components in the perturbations, since they are anyway dynamically decoupled from the scalar modes at the considered perturbative order and we only focus on the study of this latter type of perturbations~\cite{Bardeen80}.

\section{The Mukhanov-Sasaki variable}\label{ap:gauge-inv}

This Appendix provides the relation between the Mukhanov-Sasaki variable and the perturbations of the matter field in the longitudinal gauge.

The Mukhanov-Sasaki variable, together with its canonically conjugate momentum, was given in Ref.~\cite{fmo2} in terms of the scaled perturbation of the matter field of our model. In the classical theory, and in the longitudinal gauge, the two descriptions are related by the following canonical transformation of the inhomogeneous sector:
\begin{align}\label{eq_MS-potent}
v_{\vec n,\epsilon} &= A_n \tilde{f}_{\vec n,\epsilon}+B_n \tilde{\pi}_{\tilde{f}_{\vec n,\epsilon}},\\
\pi_{v_{\vec n,\epsilon}} &= C_n \tilde{f}_{\vec n,\epsilon}+D_n \tilde{\pi}_{\tilde{f}_{\vec n,\epsilon}};
\end{align}
where
\begin{align}
A_{n}& = 1-\frac{8\pi G \pi_\phi^2}{\tilde\omega^2_n v^{4/3}},\\
B_{n}& = -\frac{4\pi G\gamma \pi_\phi^2}{\tilde\omega^2_n \beta v^{5/3}},\\
C_{n} &= 4\pi G\gamma\frac{\pi_\phi^2}{\beta v^{5/3}}\left[1+\frac{1}{2\tilde\omega^2_n v^{4/3}}\left(11\frac{v^2\beta^2}{\gamma^2}-20\pi G\pi_\phi^2\right)\right],\\
D_n &= 1+\frac{2\pi G\gamma^2 \pi_\phi^2}{\tilde\omega^2_n v^{10/3}\beta^2}\left(\frac{v^2\beta^2}{\gamma^2}-4\pi G \pi_\phi^2\right).
\end{align}
Although these expressions seem complicated, one can check that the corresponding Hamilton equations (in the classical theory, or alternatively neglecting the holonomy corrections) reduce to
\begin{align}
v_{\vec n,\epsilon}' &=\pi_{v_{\vec n,\epsilon}},\\
\pi_{v_{\vec n,\epsilon}}' &=-\left(\tilde{\omega}^2_n-\frac{z''}{z}\right)v_{\vec n,\epsilon},
\end{align}
with $z$ defined in Eq. \eqref{eq:ms-z}. From these expressions it is straightforward to arrive at Eq.~\eqref{eq:Mukhanov-Sasaki}.

\section{Time dependent potentials}\label{ap:potentials}

This Appendix includes details about the time dependent potentials associated to the different descriptions of the perturbations that have been studied in this work.

\subsection{Scaled matter perturbation}

Let us consider the time dependent function $V_{\tilde f_{\vec n,\epsilon}}$ multiplying $\tilde f_{\vec n,\epsilon}$ in Eq.~\eqref{eq:KG-like}, once the contribution of the Laplacian is removed. It is clearly mode dependent and has a non-trivial dependence on the background functions. In the ultraviolet limit, however, it acquires a simple form. Let us hence consider the limit $\tilde\omega_n\to\infty$, and restrict the study to the regime in which holonomy corrections can be neglected. Under these assumptions, the time dependent potential takes the form
\begin{equation}\label{eq:timedeppota}
\lim_{\tilde{\omega}_n\rightarrow\infty}V_{\tilde f_{\vec n,\epsilon}}=-\frac{1}{2v^{4/3}}\left[\frac{v^2\beta^2}{\gamma^2}+28\pi G\pi_\phi^2 \right].
\end{equation}
Moreover, on the constraint surface the factor $v^2\beta^2$ is a constant of motion up to linear order in the perturbations (consistent with our truncation scheme in what concerns the calculation of the potential) whose value is proportional to $\pi_\phi^2$. Using this fact, one can easily arrive to Eq.~\eqref{eq:timedeppot}.

\subsection{Mukhanov-Sasaki variable}

The gauge invariant $v_{\vec n,\epsilon}$, defined by the transformation~\eqref{eq_MS-potent}, satisfies Eq.~\eqref{eq:Mukhanov-Sasaki} in the regime in which the functions $\Omega$ and $\Lambda$ [introduced in Eqs.~\eqref{eq:omega} and \eqref{eq:lambda} respectively] are well approximated by $v\beta$. The corresponding time dependent potential is determined by the quotient of the second derivative of the function $z$ [defined in Eq.~\eqref{eq:ms-z}] with respect to the conformal time, and the function $z$ itself. In the studied regime, one can see that the second derivative $z''$ can be expressed as
\begin{equation}
z'' = \frac{\gamma^3\pi_\phi}{2v^4\beta^3}\left(\frac{v^2\beta^2}{\gamma^2}-4\pi G\pi_\phi^2\right)^{2}.
\end{equation}
Using the constraint (under the same approximations), it is not difficult to obtain Eq.~\eqref{eq:tdp-MS}.


\begin{thebibliography}{M}
\bibitem{LQC} M. Bojowald, Living Rev. Rel. \textbf{11}, 4 (2008);
G. A. Mena Marug\'an, AIP Conf. Proc. \textbf{1130}, 89 (2009); J. Phys. Conf. Ser. {\bf 314}, 012012 (2011); A. Ashtekar and P. Singh, Class.\ Quantum Grav.\ \textbf{28}, 213001 (2011); K. Banerjee, G. Calcagni, and M. Mart\'\i n-Benito, SIGMA \textbf{8}, 016 (2012).

\bibitem{LQG} T. Thiemann, \textit{Modern Canonical Quantum General Relativity} (Cambridge University Press, Cambridge, UK, 2007);
K. Giesel and H. Sahlmann, Proc. Sci., QGQGS 2011, 002 (2011);
A. Ashtekar and J. Lewandowski, Class. Quantum Grav. \textbf{21}, R53
(2004).

\bibitem{APS} A. Ashtekar, T. Paw\l{}owski, and P. Singh, Phys.\ Rev.\ D \textbf{73}, 124038 (2006).

\bibitem{APS2} A. Ashtekar, T. Paw\l{}owski, and P. Singh, Phys.\ Rev.\ D \textbf{74}, 084003 (2006).

\bibitem{mmo} M. Mart\'in-Benito, G. A. Mena Marug\'an, and J. Olmedo, Phys. Rev. D \textbf{80}, 104015 (2009).

\bibitem{recall} T. Paw\l{}owski, and W. Kami\'nski, Phys. Rev. D \textbf{81}, 084027 (2010).

\bibitem{reca} A possible explanation of the accuracy of the effective description of this semiclassical behavior in loop quantum cosmology is presented in C. Rovelli and E. Wilson-Ewing, arXiv:1310.8654. 

\bibitem{moreFLRW} A. Ashtekar, T. Paw\l{}owski, P. Singh, and K. Vandersloot, Phys.\ Rev.\ D \textbf{75}, 024035 (2007);
\L{}. Szulc, W. Kami\'nski, and J. Lewandowski, Class.\ Quantum Grav.\ \textbf{24}, 2621 (2007);
E. Bentivegna and T. Paw\l{}owski, Phys.\ Rev.\ D \textbf{77}, 124025 (2008);
W. Kami\'nski and T. Paw\l{}owski, Phys.\ Rev.\ D \textbf{81}, 024014 (2010);
T. Paw\l{}owski and A. Ashtekar, Phys.\ Rev.\ D \textbf{85}, 064001 (2012).

\bibitem{anis} D. W. Chiou, Phys.\ Rev.\ D \textbf{75}, 024029 (2007); M. Mart\'\i{}n-Benito,
G. A. Mena Marug\'an, and T. Paw\l{}owski, Phys.\ Rev.\ D \textbf{78}, 064008 (2008); \textbf{80}, 084038 (2009); \L{}. Szulc, Phys.\ Rev.\ D \textbf{78}, 064035 (2008); A. Ashtekar and E. Wilson-Ewing, Phys.\ Rev.\ D \textbf{79}, 083535 (2009); \textbf{80}, 123532 (2009);
E. Wilson-Ewing, Phys.\ Rev.\ D \textbf{82}, 043508 (2010).

\bibitem{gowdy} M. Mart\'\i{}n-Benito, L. J. Garay, and G. A. Mena Marug\'an, Phys.\ Rev.\ D \textbf{78}, 083516 (2008);
G. A. Mena Marug\'{a}n and M. Mart\'{\i}n-Benito,  Int. J. Mod. Phys. A \textbf{24}, 2820 (2009);
L. J. Garay, M. Mart\'\i{}n-Benito, and G. A. Mena Marug\'an, Phys.\ Rev.\ D \textbf{82}, 044048 (2010);
M. Mart\'\i{}n-Benito, G. A. Mena Marug\'an, and E. Wilson-Ewing, Phys.\ Rev.\ D \textbf{82}, 084012 (2010);
M. Mart\'\i{}n-Beni\-to, D. Mart\'\i{}n-de Blas, and G. A. Mena Marug\'an, Phys.\ Rev.\ D \textbf{83}, 084050 (2011).

\bibitem{inflation} A.~R.~Liddle and D.~H.~Lyth, \textit{Cosmological Inflation and Large-Scale Structure} (Cambridge University Press,
Cambridge, England, 2000); J. Martin, Lect.\ Notes Phys.\ \textbf{669}, 199 (2005).

\bibitem{effecti} V. Taveras, Phys. Rev. D \textbf{78}, 064072 (2008).

\bibitem{sloan} A. Ashtekar and D. Sloan, Phys. Lett. B {\bf 694}, 108 (2010); Gen. Rel. Grav. \textbf{43}, 3619 (2011).

\bibitem{miel} D. Langlois, arXiv:1001.5259.

\bibitem{fmo} M. Fern\'andez-M\'endez, G. A. Mena Marug\'an, J. Olmedo, and J. M.  Velhinho, Phys. Rev. D \textbf{85}, 103525 (2012).

\bibitem{fmo1} M. Fern\'andez-M\'endez, G. A. Mena Marug\'an, and J. Olmedo, Phys. Rev. D \textbf{86}, 024003 (2012).

\bibitem{fmo2} M. Fern\'andez-M\'endez, G. A. Mena Marug\'an, and J. Olmedo, Phys. Rev. D \textbf{88}, 044013 (2013).

\bibitem{AAN1} I. Agullo, A. Ashtekar, and W. Nelson, Phys.\ Rev.\ Lett.\ \textbf{109}, 251301 (2012); Class. Quantum Grav. \textbf{30}, 085014 (2013).

\bibitem{AAN2} I. Agullo, A. Ashtekar, and W. Nelson, Phys.\ Rev.\ D \textbf{87}, 043507 (2013).

\bibitem{corralg} T. Cailleteau, L. Linsefors, and A. Barrau, arXiv:1307.5238;  T. Cailleteau, A. Barrau, J. Grain, and F. Vidotto, Phys. Rev. D 86, 087301 (2012). 

\bibitem{LCBG13} L. Linsefors, T. Cailleteau, A. Barrau, and J. Grain, Phys. Rev. D \textbf{87}, 107503 (2013).

\bibitem{BCGM14} A. Barrau, T. Cailleteau, J. Grain, and J. Mielczarek, arXiv:1309.6896. 

\bibitem{Wilson-Ewing12} E. Wilson-Ewing, Class. Quant. Grav. \textbf{29}, 215013 (2012).

\bibitem{bmp1} D. Brizuela, G. A. Mena Marug\'an, and T. Paw{\l}owski, \textbf{27}, 052001 (2010).

\bibitem{bmp} D. Brizuela, G. A. Mena Marug\'an, and T. Paw{\l}owski, Phys. Rev. D \textbf{84}, 124017 (2011).

\bibitem{MSvariable} M. Sasaki, Prog.\ Theor.\ Phys.\ \textbf{76}, 1036 (1986); V. Mukhanov, Sov.\ Phys.\ JETP \textbf{67}, 1297 (1988).

\bibitem{mukhanov} V. Mukhanov, \textit{Physical Foundations of Cosmology} (Cambridge University Press, Cambridge, UK, 2005).

\bibitem{verner} J. H. Verner, SIAM Journal on Numerical Analysis \textbf{15}, 772 (1978).

\bibitem{bounce} Y. F. Cai, D. A. Easson, and R. Brandenberger, J. Cosm. Astropart. Phys. \textbf{08} (2012) 020.

\bibitem{Bardeen80} J. M. Bardeen, Phys.\ Rev.\ D \textbf{22}, 1882 (1980).

\end{thebibliography}
\end{document}